\documentclass[final,5p,times,twocolumn,authoryear]{elsarticle}

\usepackage{amssymb}

\usepackage{amsmath}
\usepackage{bm}
\usepackage{listings}
\usepackage{url}
\usepackage{color}

\definecolor{keywords}{RGB}{255,0,90}
\definecolor{comments}{RGB}{0,0,113}
\definecolor{red}{RGB}{160,0,0}
\definecolor{green}{RGB}{0,150,0}
 
\newcommand*{\leopy}{LEO-Py}
\newcommand{\figuresource}[1]{The figure can be recreated with the #1 script that is part of the source code of LEO-Py.}

\journal{Astronomy and Computing}

\begin{document}

\begin{frontmatter}


\title{LEO-Py: Estimating likelihoods for correlated, censored, and uncertain data with given marginal distributions}
\author{R. Feldmann}
\ead{feldmann@physik.uzh.ch}
\ead[url]{www.ics.uzh.ch/~feldmann}
\address{Institute for Computational Science, University of Zurich, Winterthurerstrasse 190, CH-8057 Zurich, Switzerland}

\title{}


\author{}

\address{}

\begin{abstract}
Data with uncertain, missing, censored, and correlated values are commonplace in many research fields including astronomy. Unfortunately, such data are often treated in an ad hoc way in the astronomical literature potentially resulting in inconsistent parameter estimates. Furthermore, in a realistic setting, the variables of interest or their errors may have non-normal distributions which complicates the modeling.
I present a novel approach to compute the likelihood function for such data sets. This approach employs Gaussian copulas to decouple the correlation structure of variables and their marginal distributions resulting in a flexible method to compute likelihood functions of data in the presence of measurement uncertainty, censoring, and missing data. I demonstrate its use by determining the slope and intrinsic scatter of the star forming sequence of nearby galaxies from observational data. The outlined algorithm is implemented as the flexible, easy-to-use, open-source Python package LEO-Py.
\end{abstract}


\end{frontmatter}


\section{Introduction}
\label{sect:Introduction}

A frequent goal of scientific data analysis is to constrain the parameters of an underlying statistical model. In many cases, the available data is incomplete, uncertain, and correlated. Simple techniques that are frequently used in the astronomical literature, such as ordinary least squares regression, can become significantly biased and even statistically inconsistent given such realistic data sets (e.g., \citealt{Tellinghuisen2000}). This highlights the need for a flexible and easy-to-apply method to carrying out statistical inference on parametric models given data with missing entries, outliers, censored data, measurement errors, and correlations.

A large body of statistics literature deals with individual aspects of such realistic data sets. For instance, a major advance in modeling correlation structures in multivariate probability distribution was the invention \citep{Sklar1959}, and subsequent development, of copula theory (e.g., \citealt{Nelsen1995, Frees1998, Kolev2006}). Data censoring is a main study object of survival analysis \citep{Kalbfleisch2002}, perhaps one of the oldest fields in statistics \citep{Aalen2010}. The treatment of missing data is likewise its own field of research in statistics \citep{Rubin1976, Schafer2002, Little2019}. The aim of the present work is to combine existing approaches into a coherent framework to enable researchers to carry out a proper analysis of realistic data sets. 

\begin{figure}[t]
\includegraphics[width=85mm]{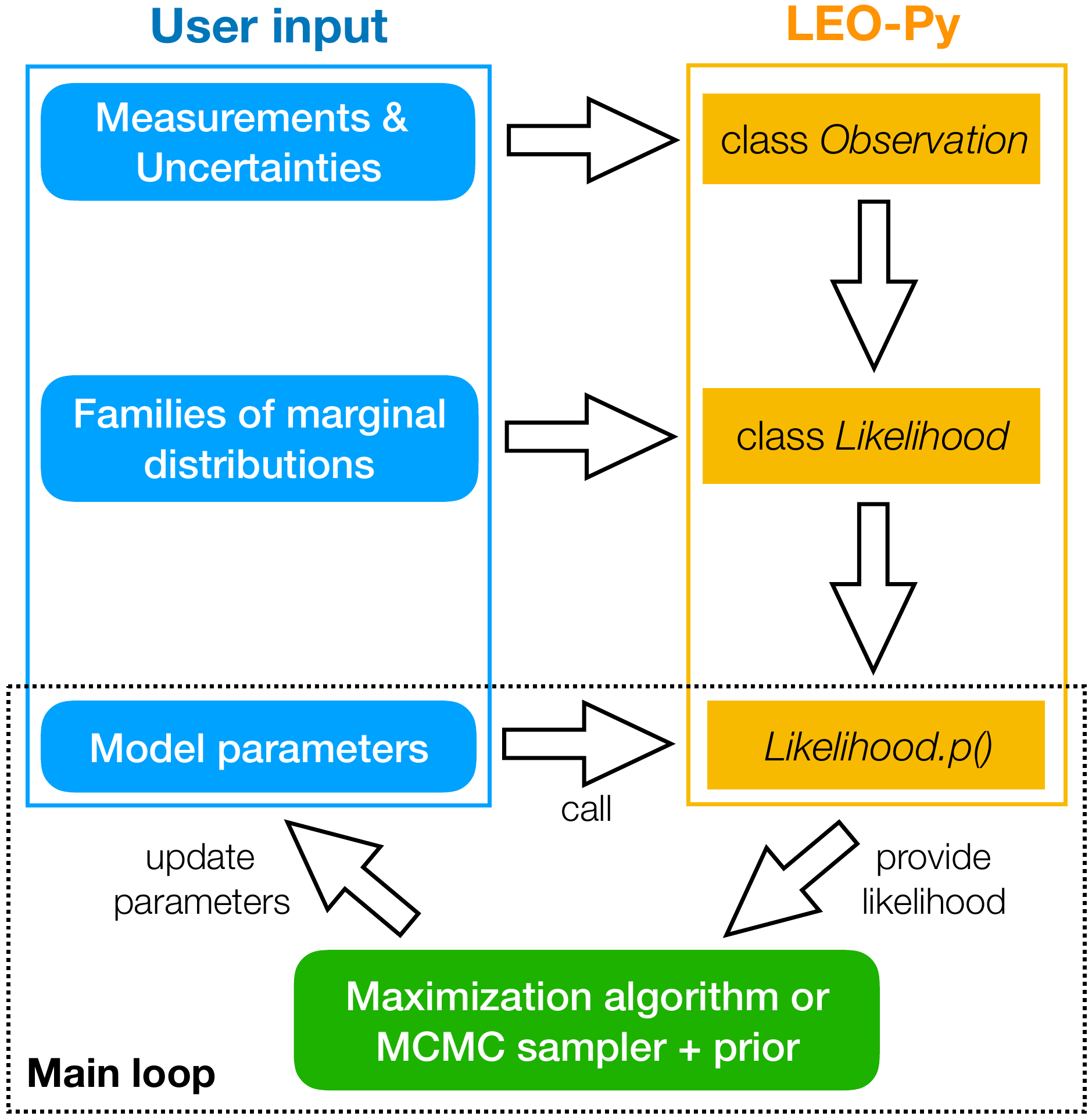}
\caption{Flowchart illustrating the main use case of the method outlined in this paper and its implementation \leopy{}. First, the user provides the observational data (potentially with missing and censored values) and their measurement uncertainties (including their correlations). Next, the user specifies the parametric families of the marginal distributions of the true (error-free) variables of interest. Now, the user may choose a set of parameters that determine the specific marginal distributions of the true variables of interest and their correlations. \leopy{} then calculates, for each such set of parameters, the likelihood of the provided set of parameters based on the given observational data. The user may employ an optimization routine, such as scipy.optimize.minimize(), to find the parameters with the highest likelihood. Alternatively, given a prior on the parameters and a Markov Chain Monte Carlo sampler, such as emcee \citep{Foreman-Mackey2012a}, the user can sample the parameter space to compute their posterior distribution.}
\label{fig:flowchart}
\end{figure}

At its basic level, the framework outlined in this paper computes the likelihood of the model parameters given uncertain, incomplete, and correlated data. This likelihood function can then be used, e.g., to find the most likely parameters via a likelihood maximization or to determine the posterior parameter distribution via Bayesian inference. Fig.~\ref{fig:flowchart} illustrates the purpose and main application of the method presented in this paper.

The method outlined in this work contains several novel features. First, it offers an efficient method to compute the correlation between observed variables (those with measurement errors) given the correlation between the true (latent) variables and the correlated measurement errors. In particular, the true variables and the measurement errors need not have multivariate normal distributions but can have arbitrary marginal, continuous probability distributions. Second, it allows a data set to contain both missing, censored, correlated, and uncertain data \emph{at the same time}. Third, while many software tools in statistics are written as R packages, Python is arguably the more commonly used computer language for data analysis in astronomy and astrophysics. The method presented in this work has been implemented as a flexible, easy-to-use, and open source Python package.

Aside from the literature mentioned above, the present work also builds on previous publications in the astro-statistical community. \cite{Kelly2007} developed a general framework to deal with realistic data sets in the regression setting. Many of the challenges mentioned above, such as data incompleteness and uncertainty, are addressed there. However, the approach described by \cite{Kelly2007}  is specific to the regression setting, only allows a single dependent variable, and furthermore, models the independent variables as Gaussian mixtures. The approach presented here generalizes \cite{Kelly2007} (i) by enabling any number of dependent and independent variables, (ii) by allowing a complex correlation structure between them, and (iii) by modeling the marginal distributions of the variables as arbitrary user-defined, continuous probability distributions. \cite{Robotham2015} introduced a multi-dimensional linear regression method (and an associated R package) that handles correlated measurement uncertainties and intrinsic scatter. A strong limitation of their method is that it performs `orthogonal regression' which is only fully consistent if the data distribution along the regression line is uniform \citep{Hogg2010}. In contrast, the method introduced here, does not suffer from this limitation.  Finally, the present work has been heavily influenced by the pedagogic summaries of parametric fitting and astro-statistical methods provided by \cite{Hogg2010}, \cite{Feigelson2012}, and \cite{Loredo2012}.

This paper is structured as follows. Section \ref{sect:Motivation} highlights the basic challenges when dealing with realistic data sets. The core methodology of this paper is subsequently presented in section \ref{sect:Method}. Section \ref{sect:Code} contains a short description of how to use the Python implementation of the present approach to compute the likelihood in various scenarios. This 
 `Likelihood Estimation for Observational data with Python` (\leopy{}) implementation is then applied to an astrophysical problem in section \ref{sect:Application}. Specifically, this section studies the relation between star formation and stellar mass of galaxies for a data set containing missing, censored, and uncertain data. Section \ref{sect:Summary} summarizes the main results and concludes the paper.

\section{Motivation}
\label{sect:Motivation}

In this section, I will show a few examples of two-dimensional data sets with missing, censored, uncertain and/or correlated entries and illustrate how well model parameters can be recovered if the likelihood is known in parametric form. For simplicity of presentation, I will not conduct a full Bayesian computation of the multivariate posterior distribution but rather calculate point estimates of the most likely values of the model parameters from a maximum likelihood analysis.

\begin{figure*}[t]
\begin{tabular}{cc}
\includegraphics[width=88mm]{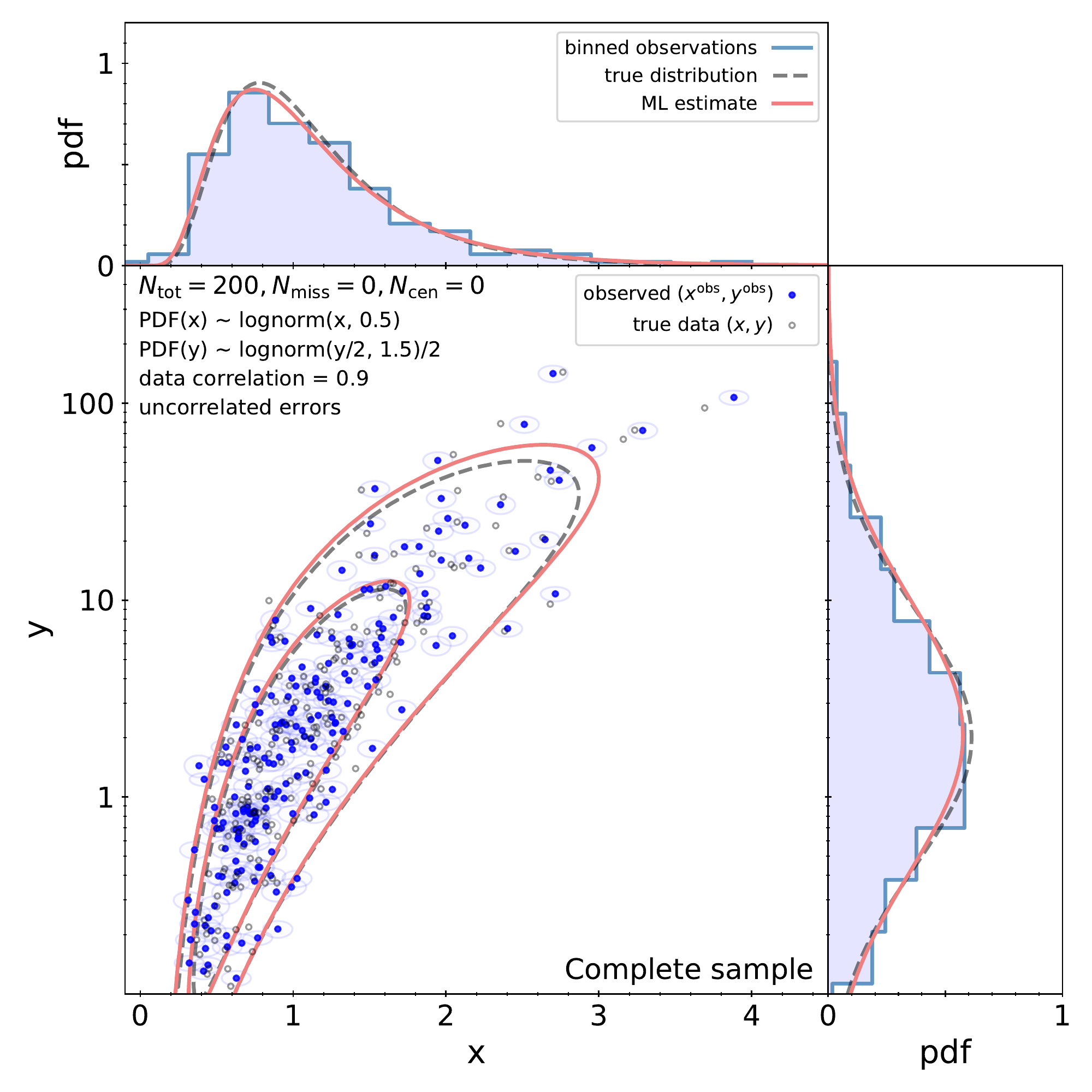} & \includegraphics[width=88mm]{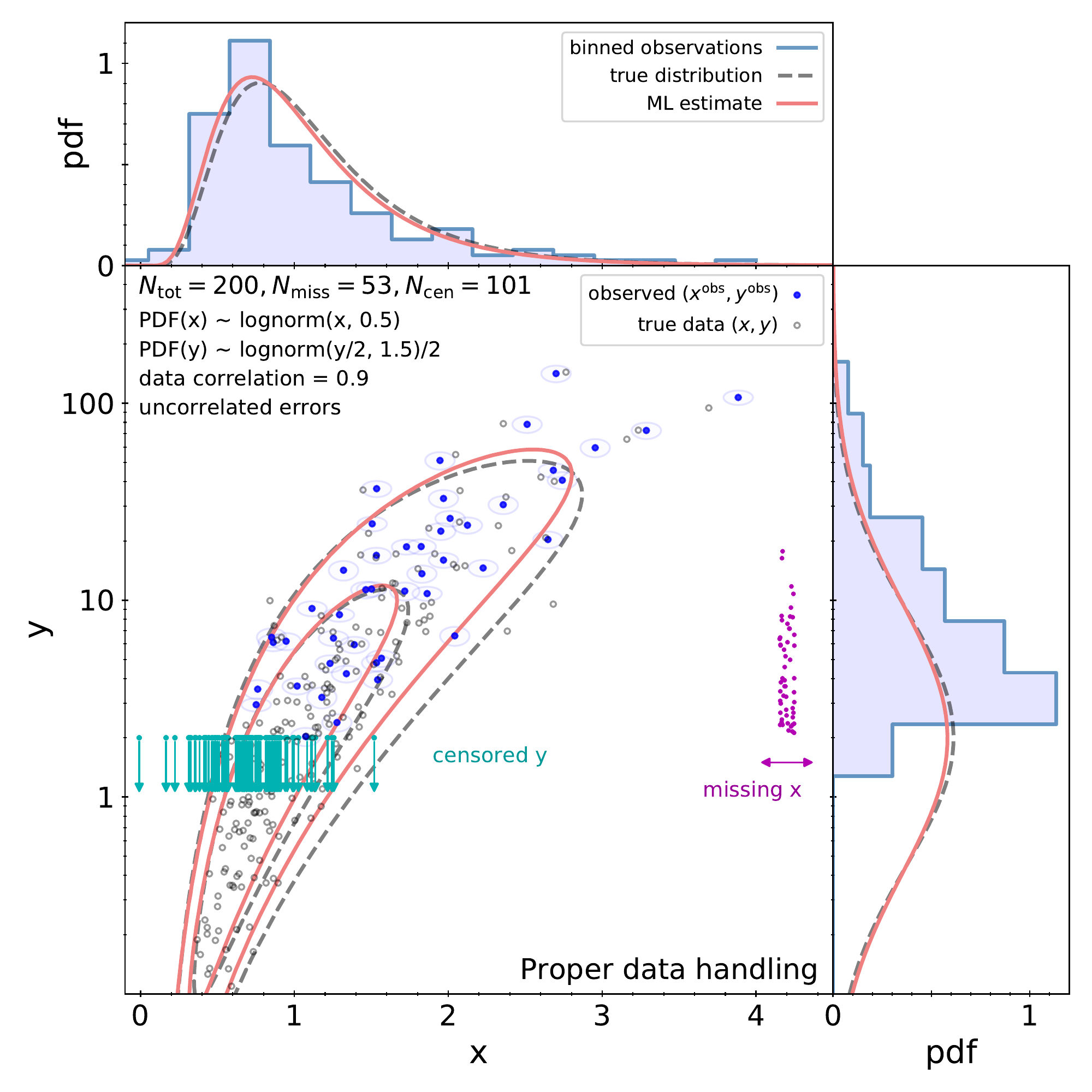}
\end{tabular}
\caption{Inferring model parameters from data with missing entries, censoring, and correlations. \figuresource{joint\_probability.py}
(Left) A random sample of 200 i.i.d. data points from a joint distribution with log-normal marginal distributions and with a correlation captured by a Gaussian copula. The parameters of the joint probability distribution are provided in the legend of the figure. Note that the $x$ axis is linear while the $y$ axis is on a logarithmic scale. (Right) The same random sample but after censoring all 101 observations falling below $y^{\rm obs}=2$ and after masking the $x^{\rm obs}$ values of 53 of the remaining observations stochastically based on their value of $y^{\rm obs}$. 
In each panel, a scatter plot shows the true data ($x$ and $y$ as empty circles) and the observed data ($x^{\rm obs}$ and $y^{\rm obs}$ as filled circles). Contour lines show the joint pdf containing $68\%$ and $95\%$ of the overall probability of the true distribution (dashed lines) and the ML estimate of the true distribution (solid lines). The marginal pdfs and the empirical densities of the data are shown at the top and the right of each panel, respectively.
In the right panel, about three quarter of the 200 individual observations are censored (cyan arrows) or missing (small filled circles).
Nevertheless, a proper treatment of data censoring, missing data, and data correlations allows to recover the underlying parameters of the true distribution to a reasonable degree of accuracy, see Table~\protect\ref{tab:example}.}
\label{fig:example_joint_proper}
\end{figure*}

\begin{figure*}
\begin{tabular}{cc}
\includegraphics[width=88mm]{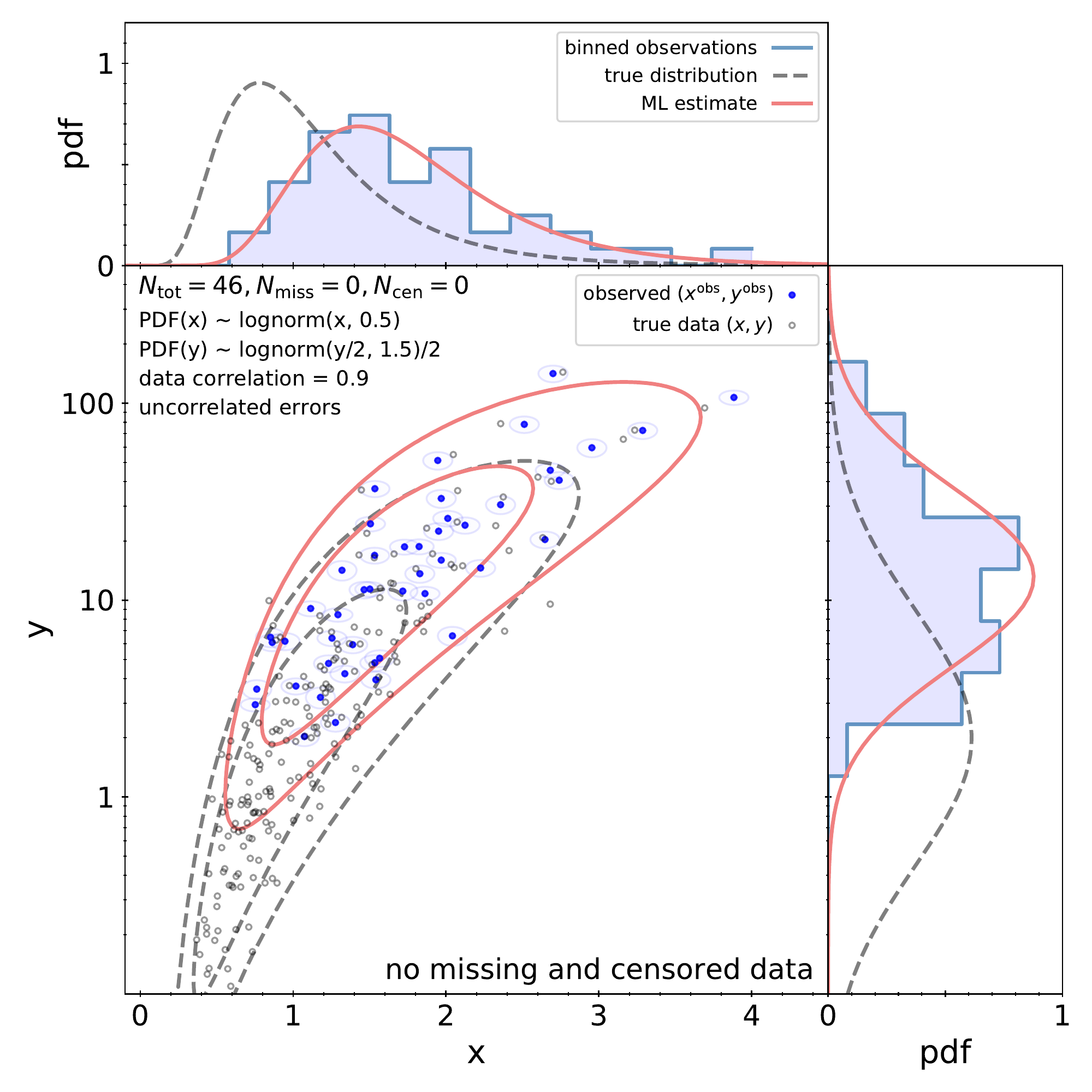} & \includegraphics[width=88mm]{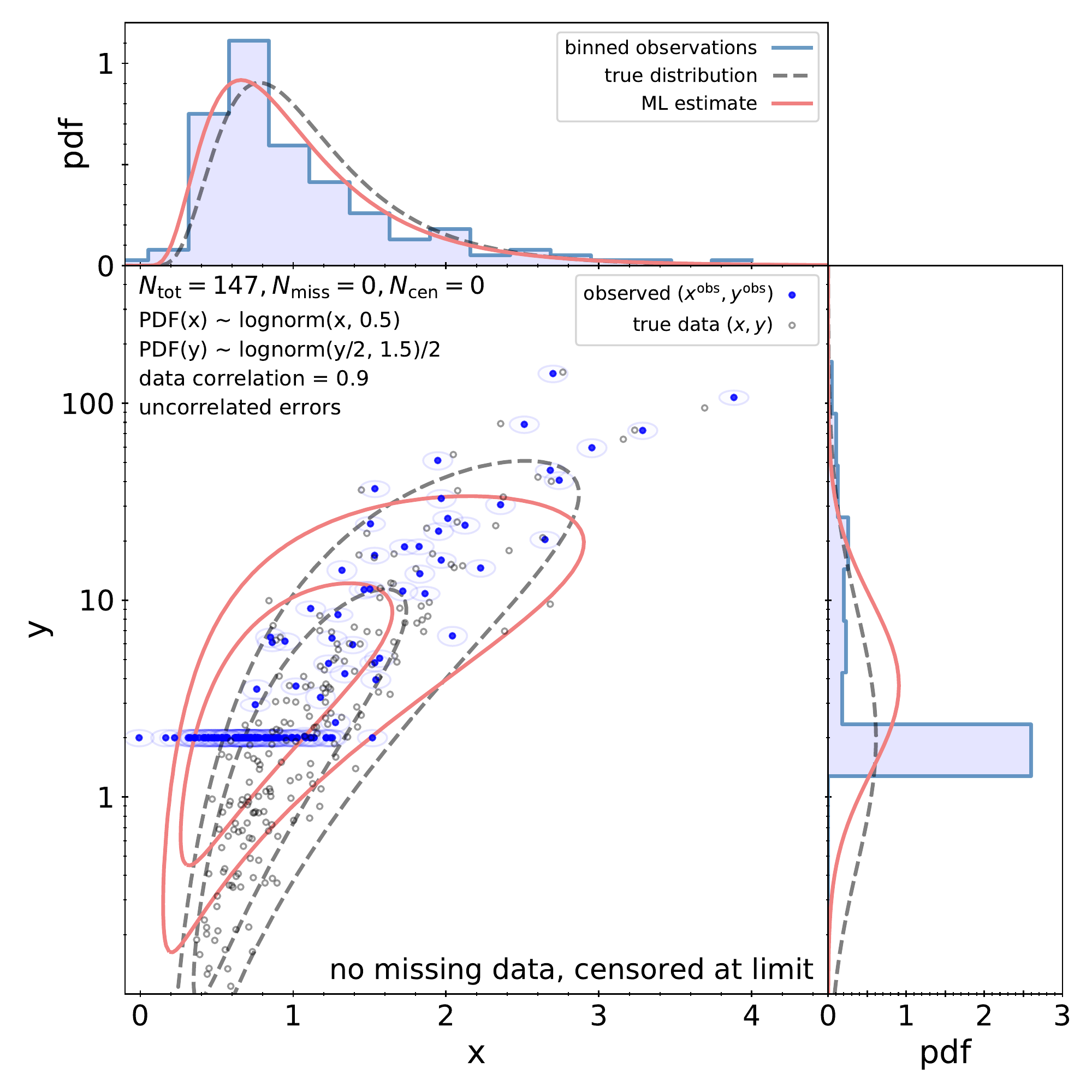} \\
\includegraphics[width=88mm]{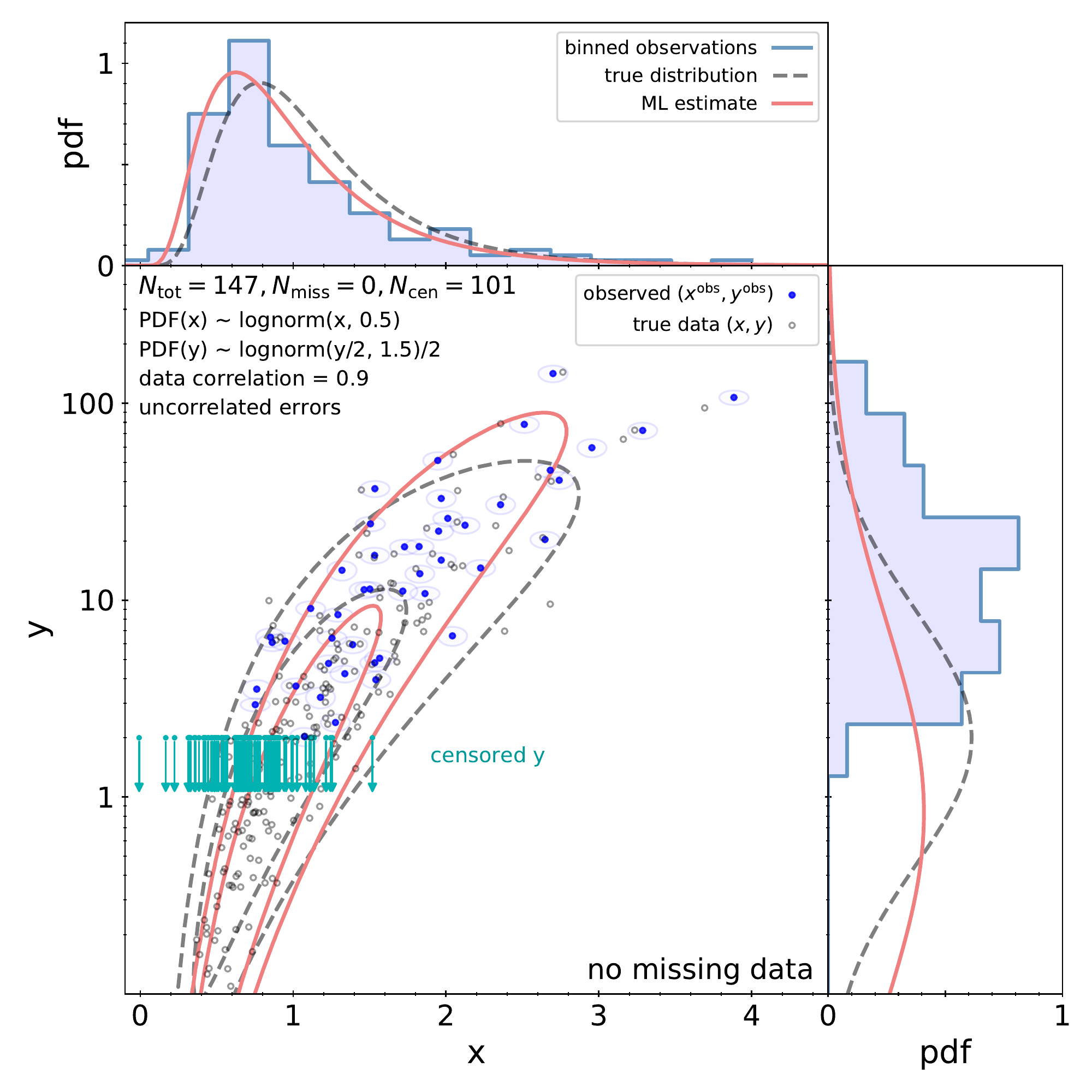} & \includegraphics[width=88mm]{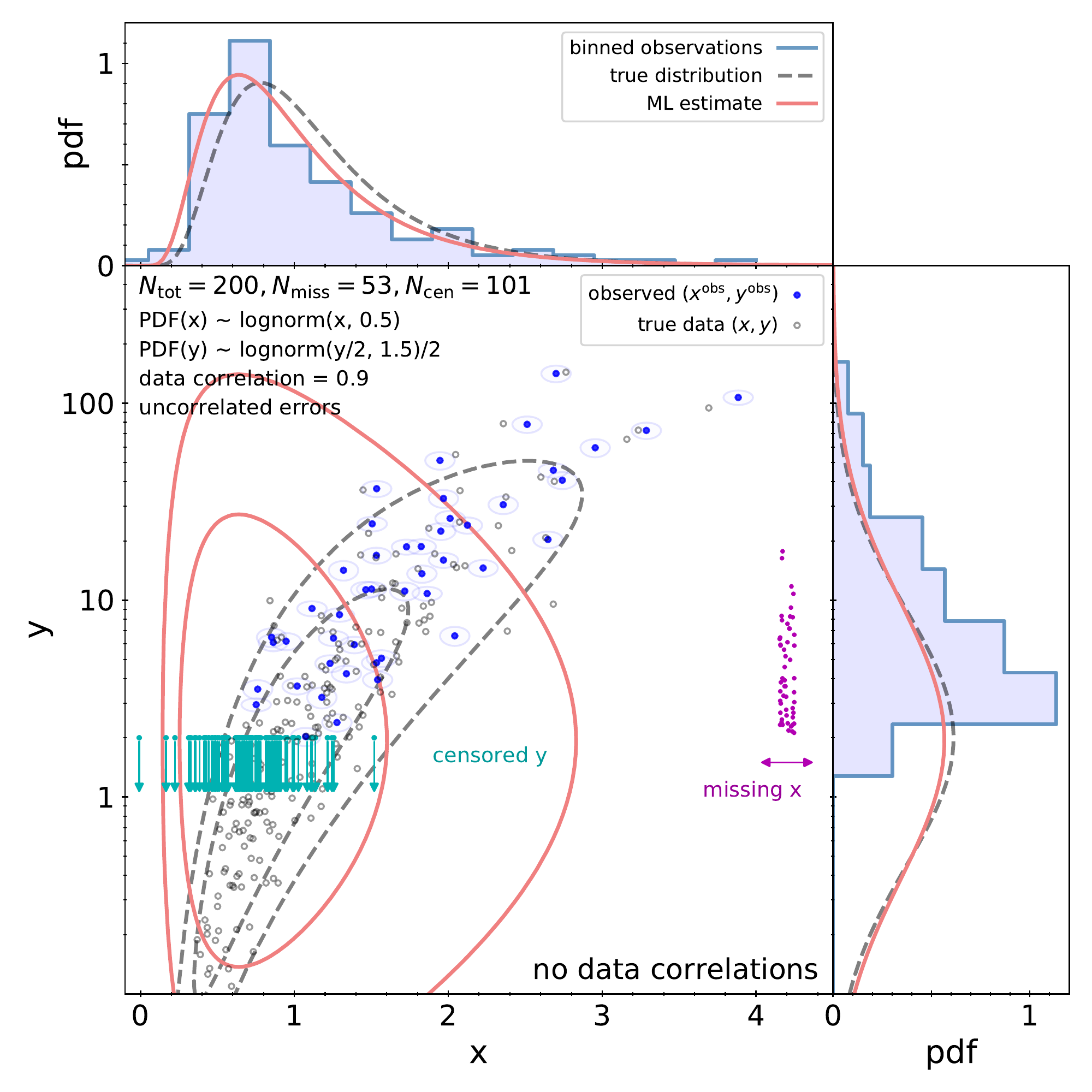} \\
\end{tabular}
\caption{Problematic strategies of inferring the underlying model parameters from data with missing entries, censoring, and correlations. The meaning of the symbols and lines is the same as in Fig.~\protect\ref{fig:example_joint_proper}. \figuresource{joint\_probability.py}
(Top left) Data points with missing or censored values are dropped from the analysis. A maximum likelihood (ML) estimate predicts model parameters of a distribution that matches well the empirical distribution of the observed data. However, these inferred model parameters are very different from the desired parameters of the true data distribution (compare solid and dashed lines). Inferences based on such an analysis are thus heavily biased. (Top right) Data with missing values are dropped and censored values are replaced by their censoring limits. While this approach performs somewhat better than ignoring censored data outright, ML estimates of the marginal and joint distributions are still heavily biased. (Bottom left) Censored data are handled properly but data with missing values are dropped. As the data are not 'missing completely at random', ignoring missing entries results in a significant bias in the parameter estimates. (Bottom right) Censored and missing data are handled properly but data correlations are ignored. This ansatz results in marginal distributions that are only mildly biased. However, the recovered joint pdf is incorrect.}
\label{fig:example_joint_problematic}
\end{figure*}

\begin{table*}
\begin{tabular}{l|ll|llllll}
\hline\hline
Parameter & Variable & true value & \multicolumn{6}{c}{Maximum likelihood estimates} \\
                  &               &                  & complete & \textbf{proper} & no miss., no cen. & no miss, cen. limit & no miss. & no corr. \\ \hline
scale ($\sigma$)    & x  & 1      &  0.989  & \textbf{0.947}  & 1.65    & 0.905  & 0.863  & 0.878 \\
                  & y  & 2     &  2.06     &  \textbf{1.99}    & 13.1   & 3.68    & 0.830   & 1.93 \\
shape ($s$)       & x  & 0.5  &  0.534   &  \textbf{0.516}  & 0.378  & 0.563  & 0.569   & 0.565 \\
                  & y & 1.5   &  1.60    &  \textbf{1.58}     & 1.05    & 1.01    & 2.25     & 1.62 \\
correlation & x,y & 0.9 & 0.908   &  \textbf{0.931}  & 0.843   & 0.797  & 0.964   & N/A \\
\hline\hline
\end{tabular}
\caption{Maximum likelihood analysis of the model parameters for the examples shown in Figs.~\protect\ref{fig:example_joint_proper} and ~\protect\ref{fig:example_joint_problematic}. Both $x$ and $y$ have lognormal distributions, $f(x, s, \sigma) = {\rm lognorm}(x/\sigma, s)/\sigma$, with  ${\rm lognorm}(x, s)=\exp\left(-\ln^2(x)/(2s^2)\right)/(x s \sqrt{2\pi})$, and $s$, $\sigma$ being the shape and scale parameters. The correlation between $x$ and $y$ is provided by a Gaussian copula. The 3rd column shows the true value of the model parameters, column 4 shows the maximum likelihood estimates based on a random sample of 200 individual observations with measurement uncertainty drawn from the joint distribution (`complete sample'). Columns 5--9 are maximum likelihood estimates based on an `incomplete' sample to highlight the importance of properly accounting for censored, missing, and correlated data. Specifically, this incomplete sample is created by censoring or marking as missing (see text) three quarters of the data in the complete sample. Column 5 (bold font) shows the result when properly analyzing the incomplete data set demonstrating that model parameters can be recovered almost as well as for the complete sample. Column 6 uses only complete cases (i.e., it ignores partially missing or censored observations). Column 7 drops missing observations but attempts to account for censored data by setting their values to the detection limits. Column 8 uses censoring information appropriately but drops observations with missing values. Finally, column 9 is the same as column 5 but assumes that $x$ and $y$ are uncorrelated.}
\label{tab:example}
\end{table*}

Fig.~\ref{fig:example_joint_proper} shows a two-dimensional data set with censored and missing data. The true data (the random variates $x$ and $y$) are drawn from a bivariate distribution with \emph{log-normal} marginal distributions and with a Gaussian copula. The parameters of the bivariate distribution are provided in the 3rd column of Table~\ref{tab:example}.

Neither $x$ nor $y$ is directly accessible. Instead, the actually measured data values ($x^{\rm obs}$ and $y^{\rm obs}$) are derived from $x$ and $y$ by adding random variates drawn from a joint normal distribution with a diagonal covariance matrix, i.e., the errors are uncorrelated in this example. A set of 200 observations created this way forms a `complete sample' as it does not feature missing or censored data. A second data set is created from the former sample by marking a fraction of the $x^{\rm obs}$ entries as missing (i.e., the data entries are replaced by `nan`) with a probability that depends on $y^{\rm obs}$. Also, in this second sample, $y^{\rm obs}$ values below 2 are marked as censored, i.e., $y^{\rm obs}$ is discarded but an upper  limit of 2 is recorded.

Given both samples, the goal is to estimate the parameters of the underlying bivariate distribution of $x$ and $y$.  The following point estimates of the distribution parameters are obtained via a likelihood maximization with the help of \leopy{} to compute the likelihood and with the basinhopping routine \citep{Wales1997} and the SLSQP minimizer to find the global minimum \citep{Kraft1994}. Both the basinhopping routine and the SLSQP minimizer are part of the scipy.optimize module that is part of the SciPy package \citep{Jones2001}. The analysis below can be reproduced with a Python script that is included in the \leopy{} package.

The left panel of Fig.~\ref{fig:example_joint_proper} shows the recovery of the distribution parameters using the complete sample. This result serves a baseline to assess the recovery based on the second, more challenging, data set. Recovered distribution parameters are listed in column 4 of Table~\ref{tab:example}. They differ from the true parameters by only a few percent.

The right panel of Fig.~\ref{fig:example_joint_proper} shows that the parameters of the joint $x, y$ distribution can be recovered without significant bias even with a large fraction of the data missing or censored. The recovered distribution parameters are listed in column 5 of Table~\ref{tab:example}. They differ by less than 5\% from the parameter estimates based on the complete sample.

Missing and censored data are not always properly handled in the astronomical and astrophysical literature. To illustrate the potential consequences, I show in Fig.~\ref{fig:example_joint_problematic} the maximum likelihood estimates for various problematic choices of dealing with incomplete data sets. The estimated parameters are also provided in Table~\ref{tab:example}. 

The top left figure shows the result of keeping only complete data set, i.e., dropping any observations with missing or censored values. In this case, the likelihood maximization is not able to recover the underlying distribution of the true data ($x, y$). Instead it is strongly biased towards fitting the distribution of the uncensored and non-missing observed data ($x^{\rm obs}$, $y^{\rm obs}$). Hence, inferences based on analyses that ignore censored and missing data may be incorrect. 

Another strategy frequently used in the literature is to substitute detection limits for the censored values. As the top right panel of Fig.~\ref{fig:example_joint_problematic} shows, the distribution of the true data ($x, y$) is still not recovered very well. The bottom left panel of Fig.~\ref{fig:example_joint_problematic} shows the result of treating censored data properly but ignoring missing data. As the missing $x^{\rm obs}$ values depend on $y^{\rm obs}$, the data is 'missing at random' (MAR), but \emph{not} 'missing completely at random' (MCAR). Hence, dropping observations with such missing data can make parameter estimates statistically inconsistent. Finally, the bottom right panel of the same figure shows the outcome of ignoring data correlations. While the marginal distributions are approximately recovered, the lack of a data correlation implies that the predicted joint distribution is incorrect. Missing and censored data, as well as data correlations need to be handled properly to avoid large biases and inconsistencies when inferring the underlying model parameters.

\begin{figure*}
\begin{tabular}{cc}
\includegraphics[width=85mm]{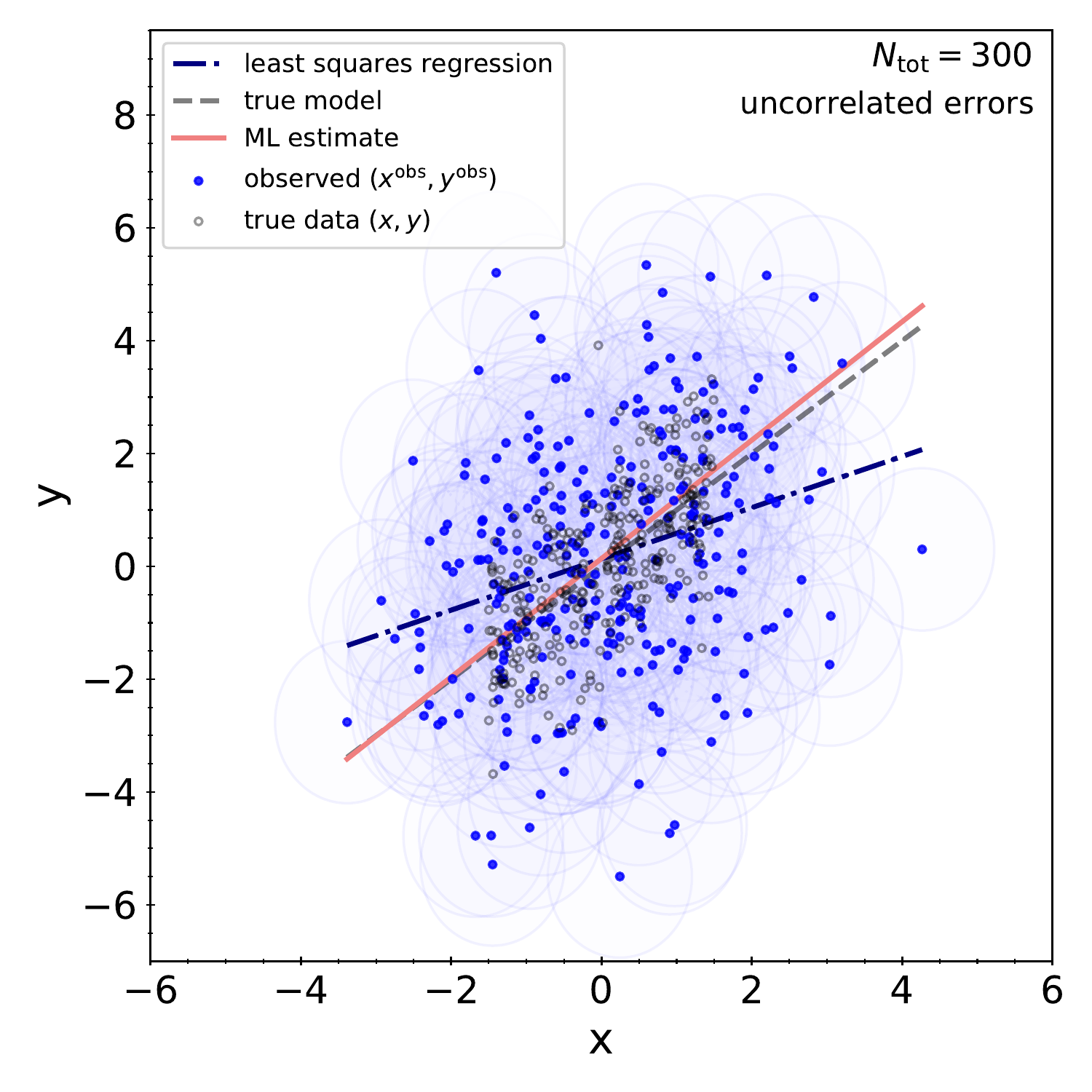} & \includegraphics[width=85mm]{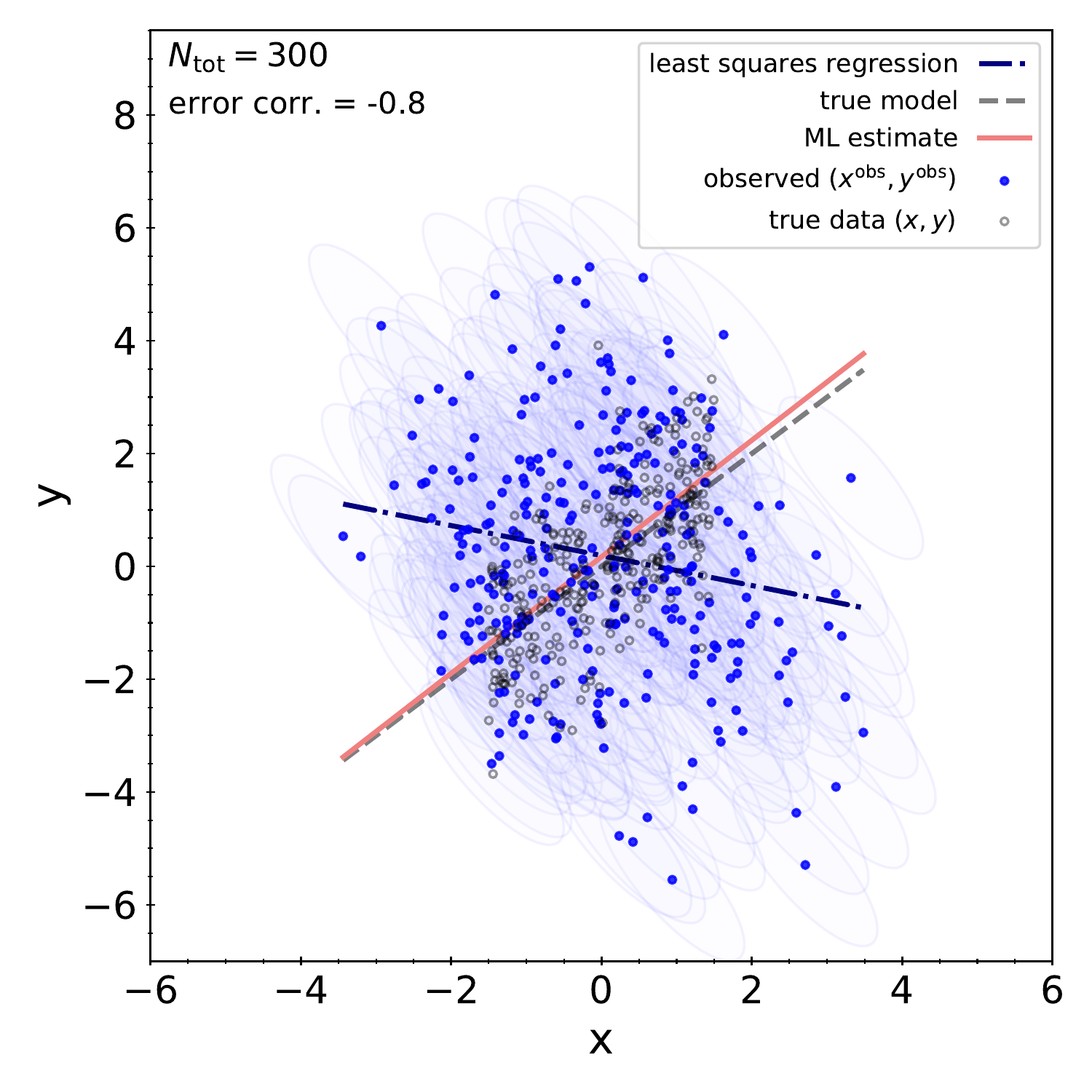}
\end{tabular}
\caption{The regression problem of constraining $y\sim{}x$ with $x$ and $y$ both subject to measurement uncertainties. \figuresource{lin\_regression.py} Measured values ($x^{\rm obs}$ and $y^{\rm obs}$) are shown as filled circles with observational uncertainties (68\% confidence contours) indicated by lightly-shaded ellipses centered on the observational data point, while true data values ($x$ and $y$) are shown as empty circles. Measurement errors are uncorrelated in the left panel and anti-correlated in the right panel.  The dashed line shows the input model $\bar{y}(x)=x$. At fixed $x$, the true $y$ values are normally distributed around $\bar{y}$ with a standard deviation of 1. The dot-dashed line shows the result of fitting $y^{\rm obs}$ as function of $x^{\rm obs}$ with ordinary least squares regression. The slope predicted by this naive approach differs significantly from the true slope and can even have the wrong sign (see panel on the right). Instead, a proper maximum likelihood optimization (solid line) that accounts for the measurement error covariance infers a slope ($1.05$ for the left panel, $1.03$ for the right panel), intercept ($0.14, 0.17$) and intrinsic scatter of $y$ given $x$ ($1.1, 1.05$) in good agreement with the input model.}
\label{fig:example_regression}
\end{figure*}

Linear regression problems of the type $y = \beta_1x + \beta_0 + \epsilon$ are common in many fields of science, including astrophysics. An often employed technique, ordinary least squares, provides the best linear unbiased estimators for slope ($\beta_1$) and intercept ($\beta_0$) if the errors $\epsilon$ of the dependent variable have an expectation value of zero, are homoscedastic, and are uncorrelated (see Gauss--Markov theorem). However, the standard assumption of regression problems such as the above is that the independent (predictor) variables are known without error \citep{Montgomery2012}. Indeed, applying the standard regression techniques to data with errors in the predictors can result in biases, such as attenuation bias (e.g., \citealt{Frost2000}). Various methods have been developed to deal with uncertainties in the predictors. A common approach of such `errors-in-variables` models is to introduce latent variables that represent the true values of the dependent and independent variables and add an error term to each to obtain the measured dependent and independent variables. I will follow this ansatz for the example below. Specifically, I will assume that
\[
(x^{\rm obs}, y^{\rm obs})^\top = (x, \beta_1x + \beta_0 + \epsilon)^\top + \eta,
\]
where $\epsilon$ is a normally distributed random variate with zero expected value and with standard deviation $\sigma$ representing intrinsic scatter of $y$, while $\eta$ is a random variate drawn from a two-dimensional joint normal distribution with a \emph{known} covariance matrix $\Sigma$.  For each observation, the likelihood of the model parameters can then be calculated as 
\begin{equation}
\begin{split}
\mathcal{L}(\beta_0, \beta_1, \sigma) &= p(x^{\rm obs}, y^{\rm obs}\vert{}\beta_0, \beta_1, \sigma) \\
&= \int p(x^{\rm obs}, y^{\rm obs}\vert{}x, \beta_0, \beta_1, \sigma)p(x)dx \\
&=\int \phi_{\tilde{\Sigma}}(x^{\rm obs} - x, y^{\rm obs} -  \beta_1x + \beta_0 ) p(x) dx,
\end{split}
\label{eq:LikeExample}
\end{equation}
where $\phi_{\tilde{\Sigma}}$ is a multivariate normal distribution with zero mean and covariance matrix $\tilde{\Sigma}$. The latter has the matrix elements $\tilde{\Sigma}_{xx}=\Sigma_{xx}$, $\tilde{\Sigma}_{xy}=\Sigma_{xy}$, and $\tilde{\Sigma}_{yy}=\Sigma_{yy}+\sigma^2$.  

Fig.~\ref{fig:example_regression} shows a data set created as described above with $\beta_1=1$, $\beta_0=0$, $x$ drawn from a uniform distribution over the interval $[-1.5, 1.5]$, and $\sigma=1$. The observational errors are uncorrelated in the left panel and they are anti-correlated ($\rho=-0.8$) in the right panel. Specifically, $\eta\sim{}N(0, \Sigma)$ with 
\begin{align*}
\Sigma&=\begin{pmatrix}1 & 0 \\ 0 & 2.25\end{pmatrix} &\textrm{ in the left panel and } \\
\Sigma&=\begin{pmatrix}1 & -1.2 \\ -1.2 & 2.25\end{pmatrix} &\textrm{ in the right panel.}
\end{align*}

As Fig.~\ref{fig:example_regression} exemplifies, naively applying ordinary least squares to the data results in attenuation bias (if measurement errors are uncorrelated) or even in a different sign of the slope (if measurement errors are anti-correlated). In contrast, by maximizing the proper likelihood function, equation (\ref{eq:LikeExample}), the slope, intercept, and scatter of the true $y\sim{}x$ relation are recovered very well.

In summary, as  the examples in this section demonstrate, it is often necessary to account simultaneously for data censoring, missing data, data correlations, as well as measurement errors and their correlations when estimating parameters of statistical models.

\section{The general method}
\label{sect:Method}

In this section, I will describe the general approach of computing the likelihood for data with censored and missing entries, data correlations, data outliers, and (correlated) measurement errors. The discussion will generalize from the bivariate case discussed in the previous section to the multivariate case of having $K$ observed variables. 

\begin{figure}
\includegraphics[width=80mm]{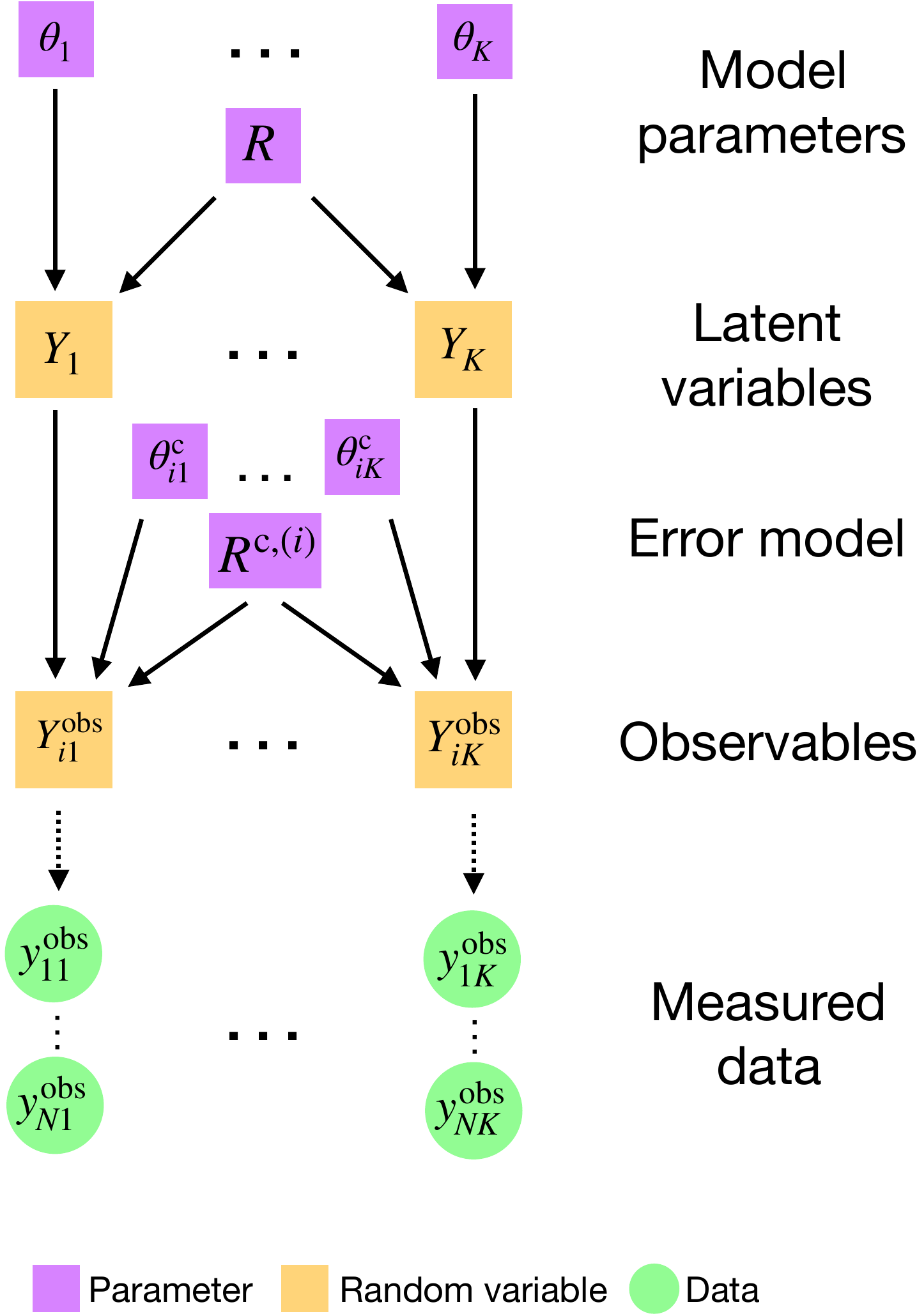}
\caption{Illustration of the multilevel model adopted in this paper. The model parameters consist of a set of parameters $\theta_j=\{\mu_j, \sigma_j, \ldots{}\}$ specifying the marginal distributions of the true (but unknown) values of observables ($\{Y_j\}$), a correlation matrix $R$ describing the correlation structure between $\{Y_j\}$, and similar parameters ($\theta^{\rm\,c}_{ij}$, $R_i^{\rm c}$) describing the error model, i.e.,~the map between the true values of observables and their measured values. Model parameters may vary with observation $i=1,\ldots{},N$ and the figure highlights the common use case of an observation-dependent error model. Large squares show random variables ($\{Y_j\}$, $\{Y_j^{\rm obs}\}$), while circles show the actually measured data (a matrix $y_{ij}^{\rm obs}$ of random variates with $1\leq{}i\leq{}N$ and $1\leq{}j\leq{}K$, where $N$ is the number of independently drawn observations and $K$ is the number of observables per observation).}
\label{fig:multilevel}
\end{figure}

Fig.~\ref{fig:multilevel} illustrates the hierarchical structure of the multilevel approach presented in this paper. $\mathbf{Y}=(Y_1,\ldots{}Y_K)$ are random variables denoting the quantities of interest, e.g., the star formation rate, the gas mass, or, to provide an example outside the astronomical context, the weight or height of a person. As mentioned before, actual measurements are \emph{not} realizations (`random variates') of $\mathbf{Y}$. Instead, each observation $i$ of the $K$ variables consists of random variates of a different set of $K$ random variables $\mathbf{Y}^{\rm obs}_{i}$ that are related to $\mathbf{Y}$ but include contributions from observation-specific measurement errors. Any two observations $i_1$ and $i_2$ with $i_1\neq{}i_2$ are assumed to be drawn independently from the distributions of their respective random variables $\mathbf{Y}^{\rm obs}_{i_1}$ and $\mathbf{Y}^{\rm obs}_{i_2}$.

For any given observation $i$, the multivariate distribution of the random variables $\mathbf{Y}^{\rm obs}_i$ is determined by two sets of input parameters that define the  `true data model` and the `error model`, respectively. Regarding the former, the parameters $\theta_j$ specify the shape of the (1-dimensional) distributions of the latent variables $Y_{j=1,\ldots{},K}$ while the correlation matrix $R$ parametrizes the correlation structure between them. The error model is given by the parameters $\theta^{\rm c}_{ij}$, and $R^{\rm c}_i$ which determine the (1-dimensional) conditional distribution of $Y^{\rm obs}_j$ given $Y_j$  as well as the correlation structure among $\mathbf{Y}^{\rm obs}$ given $\mathbf{Y}$. The parameters of the true data model and error model may vary between observations. A typical use case of the latter is to account for observation specific measurement uncertainties. 

The challenge is to compute the probability of obtaining a specific sample of random variates $\mathbf{y}^{\rm obs}_{i=1,\ldots{},N}$  drawn from the distribution of $\mathbf{Y}^{\rm obs}_{i=1,\ldots{},N}$ for a given set of model parameters. Complications include the presence of measurement uncertainty, correlations among $\mathbf{Y}$ and $\mathbf{Y}^{\rm obs}$, and problems related to data censoring and missing data.

Before describing the method in more detail, I would like to point out the main assumptions of the adopted approach. First, it only concerns itself with continuous distributions. A generalization that includes discrete or mixed continuous-discrete distributions is beyond the scope of this paper but may be studied in the future. Second, the approach is parametric in nature, i.e., distributions of $\mathbf{Y}$ and conditional distributions of $\mathbf{Y}^{\rm obs}$ given $\mathbf{Y}$ are assumed to be given in a parametric form. Finally, the correlation structure is assumed to be described by a Gaussian copula. This is a natural choice as (i) it generalizes the common case of normally distributed variables with normal error distributions, (ii) it accounts for the full range of correlation from perfect anti-correlations to perfect correlations, and (iii) lends itself to a natural parametrization and interpretation of correlations via correlation matrices. The choice of a Gaussian copula significantly simplifies the challenges that arise from data correlation, censoring, and missing data. However, other copulas might be more appropriate for specific problems.

To avoid clutter of notation, I will not indicate the dependence on the observation $i$ unless this dependence is of specific interest. Furthermore, I will use various shorthands, such as $\int^{\mathbf{z}^{\rm lo}}d\mathbf{z}^{\rm cen}$, which should be read as $\int^{z^{\rm lo}_1}dz^{\rm cen}_1 \ldots \int^{z^{\rm lo}_K}dz^{\rm cen}_K$, or $p(y)$ as shorthand for $p(Y=y)$.
Bold face is generally used to indicate a tuple in the space of variables, e.g., $\mathbf{y}^{\rm obs}=(y^{\rm obs}_1,\ldots{},y^{\rm obs}_K)$ or $\mathbf{Y}=(Y_1,\ldots{},Y_K)$. Furthermore, $p$ is used as a generic symbol to indicate probability densities or distributions. Its meaning should be apparent from context. Finally, to simplify the notation further, I do not highlight the conditional dependence of densities and distributions on the parameters of the model and the error model in sections \ref{sect:margdist}-\ref{sect:S}. However, this dependence will be made explicit in the subsequent sections.

\subsection{The marginal distributions of observables}
\label{sect:margdist}

Let the 1-dimensional probability distribution of each latent variable $Y_j$ be $F_j$ and its density be $f_j$. The error model provides the probability distribution $F^{\rm c}_j$ and density $f^{\rm c}_j$ of measuring $Y^{\rm obs}_j=y^{\rm obs}_j$ given $Y_j=y_j$. Hence, the probability density and distribution of $Y^{\rm obs}_j$ are (via the law of total probabilities):
\begin{align}
f^{\rm obs}_j(y^{\rm obs}_j) &= \int d y_j f^{\rm c}_j(y^{\rm obs}_j\,\vert\,y_j) f_j(y_j) \label{eq:fobs} \\
F^{\rm obs}_j(y^{\rm obs}_j) &= \int d y_j F^{\rm c}_j(y^{\rm obs}_j\,\vert\,y_j) f_j(y_j) \label{eq:Fobs}.
\end{align}
 
These expressions can be simplified or solved if $f$ and $f^{\rm c}$ belong to certain families of probability densities. In general, however, both equations have to be solved numerically.

\subsection{Modeling correlations between observables}
\label{sect:corrobs}

To model the multivariate probability density and the correlations among and between $\mathbf{Y}$ and $\mathbf{Y}^{\rm obs}$, I consider a Gaussian copula on the combined space $\{Y_1^{\rm obs},\ldots,Y_K^{\rm obs},Y_1,\ldots,Y_K\}$. Specifically,
\begin{multline}
\label{eq:FullJointProbability}
p(y_1^{\rm obs},\ldots,y_K^{\rm obs},y_1,\ldots,y_K) = \\ 
\begin{split}
&f^{\rm obs}_1(y_1^{\rm obs}) \times{} \ldots \times{} f^{\rm obs}_K(y_K^{\rm obs}) \times{} f_1(y_1) \times{} \ldots \times{}f_K(y_K) \\
&\times{}c_{P}(F^{\rm obs}_1(y_1^{\rm obs}),\ldots,F^{\rm obs}_1(y_K^{\rm obs}),F_1(y_1),\ldots,F_K(y_K))
\end{split}
\end{multline}
where  $f_j^{\rm obs}$ and $f_j$ are the (marginal) probability densities of $Y_j^{\rm obs}$ and $Y_j$, $F_j^{\rm obs}$ and $F_j$ are their (marginal) distribution functions, and $c_P$ is a Gaussian copula density with correlation matrix $P$.

As shown by \cite{Sklar1959} any multivariate distribution can be written as the product of 1-dimensional distributions and a copula which captures the correlation between variables.  Adopting a Gaussian copula with a correlation matrix $P$, amounts to choosing the shape of the multivariate distribution for given marginal distributions.

The Gaussian copula density with correlation matrix $P$ for $\mathbf{u}=(u_1,\ldots{},u_K)$ is defined as
\begin{equation}
\label{eq:GaussCopula}
c_P(\mathbf{u}) = \frac{\phi_P(\Phi^{-1}(u_1),\ldots,\Phi^{-1}(u_K))}{\prod_{i=1}^K\phi(\Phi^{-1}(u_i))}.
\end{equation}
Here and in the remainder of this paper, $\Phi(x)$ denotes a standard normal distribution, $\phi$ a standard normal density, and $\phi_P(\mathbf{x})$ a multivariate normal density with covariance matrix $P$.

Given the form of the Gaussian copula, it is useful to introduce standardized variables
\begin{equation}
z_j = \Phi^{-1}(F_j(y_j)) \textrm{ and } z_j^{\rm obs}=\Phi^{-1}(F_j^{\rm obs}(y^{\rm obs}_j)),
\end{equation}
and to rewrite equations (\ref{eq:FullJointProbability}) and (\ref{eq:GaussCopula}) as:
\begin{align}
\label{eq:Pyobsy}
p(\mathbf{y}^{\rm obs},\mathbf{y}) &= p(\mathbf{z}^{\rm obs},\mathbf{z})\, \prod_{j=1}^K \frac{dz^{\rm obs}_j}{dy^{\rm obs}_j} \frac{dz_j}{dy_j}\\
\label{eq:Pzobsz}
p(\mathbf{z}^{\rm obs},\mathbf{z}) &= \phi_P(\mathbf{z}^{\rm obs},\mathbf{z})
\end{align}
In other words, $(\mathbf{Z^{\rm obs}}, \mathbf{Z})$ has a multivariate normal distribution with covariance (and correlation) matrix $P$ and each of its components has a standard normal distribution.

The probability density of $\mathbf{Y}^{\rm obs}$ can be obtained by integrating $p(\mathbf{y}^{\rm obs},\mathbf{y})$ over $\mathbf{y}$:
\begin{equation}
\label{eq:pyobs}
p(\mathbf{y}^{\rm obs}) = p(\mathbf{z}^{\rm obs})  \prod_{j=1}^K \frac{dz^{\rm obs}_j}{dy^{\rm obs}_j} = 
p(\mathbf{z}^{\rm obs})  \prod_{j=1}^K \frac{f^{\rm obs}_j(y^{\rm obs}_j)}{\phi(z^{\rm obs}_j)},
\end{equation}
with
\begin{equation}
\label{eq:pzobs}
p(\mathbf{z}^{\rm obs}) =  \int d\mathbf{z}\,\phi_P(\mathbf{z}^{\rm obs}, \mathbf{z})
\end{equation}

Using block-matrix notation we can decompose the  $2K\times{}2K$ correlation matrix $P$ into 4 separate sub-matrices:
\begin{equation}
\label{eq:Pmatrix}
P = \begin{pmatrix} S & T \\ T^\top & R \end{pmatrix}.
\end{equation}
One of the basic properties of a multivariate normal distribution is that integrating out a subset of its components results in a multivariate normal distribution with a covariance matrix that is a sub-matrix of the original covariance matrix \citep{Petersen2012}. Applied to the present case, $\mathbf{Z}$ and $\mathbf{Z}^{\rm obs}$ have multivariate normal distributions with covariance matrices $R$ and $S$, respectively. 

Equation (\ref{eq:pzobs}) has thus the simple solution $p(\mathbf{z}^{\rm obs}) =  \phi_S(\mathbf{z}^{\rm obs})$
and
\begin{equation}
\label{eq:pyobs2}
p(\mathbf{y}^{\rm obs}) = 
 \phi_S(\mathbf{z}^{\rm obs})  \prod_{j=1}^K \frac{f^{\rm obs}_j(y^{\rm obs}_j)}{\phi(z^{\rm obs}_j)}.
\end{equation}
Hence, the correlations among $\mathbf{Y}^{\rm obs}$ are described by a Gaussian copula with correlation\footnote{$S$ is a correlation matrix, because, as a principal sub-matrix of $P$, it is necessarily real and symmetric, has diagonal entries of 1, and is positive semi-definite due to Cauchy's interlace theorem.} matrix $S$. Completely analogously, correlations among $\mathbf{Y}$ are described by a Gaussian copula with correlation matrix $R$.

The matrix $S$ can be computed from the provided input parameters ($R$, $R^{\rm c}$, $\theta_j$, $\theta^{\rm c}_{j}$), see section \ref{sect:S}.

\subsection{Marginalizing and conditional probabilities}

Marginalized and conditional probability densities can be calculated analytically from equation (\ref{eq:pyobs2}) as I will now show for subsets of the variables $\{Y_1^{\rm obs}, \ldots, Y_K^{\rm obs}\}$. Specifically, let $\mathbf{Y}^{\rm obs}_a=\{Y_{a_1}^{\rm obs}, \ldots, Y_{a_L}^{\rm obs}\}$ be any such subset and $\mathbf{Y}^{\rm obs}_b=\mathbf{Y}^{\rm obs}\setminus{}\mathbf{Y}^{\rm obs}_a$ be its complement.

Analogous to above, the correlations of $\mathbf{Y}^{\rm obs}_a$ ($\mathbf{Y}^{\rm obs}_b$) are captured by a Gaussian copula with the $L\times{}L$ ($K-L\times{}K-L$) correlation matrix $S_a$ ($S_b$) that is the principal sub-matrix of $S$ with rows and columns corresponding to $\mathbf{Y}^{\rm obs}_a$ ($\mathbf{Y}^{\rm obs}_b$). To simplify notation,  it is helpful to order the indices and re-label the variables such that $a_1,\ldots{},a_L$ are the first $L$ indices. In this case, the $K\times{}K$ matrix $S$ can be written in block-matrix form as
\begin{equation}
\label{eq:Smatrixdecomp}
S = \begin{pmatrix} S_a & S_c \\ S_c^\top & S_b \end{pmatrix}.
\end{equation}

As discussed above, marginalizing the probability density of a multivariate normal distribution is straightforward. Applying this idea to equation (\ref{eq:pyobs2}) leads to 
\begin{equation}
\begin{split}
p(\mathbf{z}^{\rm obs}_a) & = \int d\mathbf{z}^{\rm obs}_{b}\phi_S(\mathbf{z}^{\rm obs}) \\
& = \phi_{S_a}(\mathbf{z}^{\rm obs}_a),
\end{split}
\end{equation}
and, thus,
\begin{equation}
p(\mathbf{y}^{\rm obs}_a) = \phi_{S_a}(\mathbf{z}^{\rm obs}_a)  \prod_{j=1}^L \frac{f^{\rm obs}_{a_j}(y^{\rm obs}_{a_j})}{\phi(z^{\rm obs}_{a_j})}.
\end{equation}

For a multivariate normal distribution of vector  $\mathbf{Z}^{\rm obs}$ with zero mean and with a covariance matrix $S$ given by (\ref{eq:Smatrixdecomp}), the conditional probability of $\mathbf{Z}^{\rm obs}_a$ given $\mathbf{Z}^{\rm obs}_b$ is again multivariate normal \citep{Petersen2012}
\begin{equation}
\label{eq:CondProbS}
p(\mathbf{z}^{\rm obs}_a\vert\mathbf{z}^{\rm obs}_b) = \phi_{S_a - S_cS_b^{-1}S_c^\top}(\mathbf{z}^{\rm obs}_a - S_cS_b^{-1} \mathbf{z}^{\rm obs}_b).
\end{equation}
The conditional probability density of $\mathbf{Y}^{\rm obs}_a$ given $\mathbf{Y}^{\rm obs}_b$ is thus
\begin{equation}
p(\mathbf{y}^{\rm obs}_a\vert\mathbf{y}^{\rm obs}_b) = \phi_{S_a - S_cS_b^{-1}S_c^\top}(\mathbf{z}^{\rm obs}_a - S_cS_b^{-1} \mathbf{z}^{\rm obs}_b) \prod_{j=1}^L \frac{f^{\rm obs}_{a_j}(y^{\rm obs}_{a_j})}{\phi(z^{\rm obs}_{a_j})}.
\end{equation}

\subsection{Computing the correlation matrix for $\mathbf{Z}^{\rm obs}$}
\label{sect:S}

As shown in section \ref{sect:margdist}, the standardized variable $\mathbf{Z}^{\rm obs}$ has a multivariate normal distribution with covariance (and correlation) matrix $S$. Similarly, $(\mathbf{Z}^{\rm obs}, \mathbf{Z})$ is multivariate normal with covariance and correlation matrix $P$. I will now show how $S$ can be derived provided the correlation matrices $R$, $R^{\rm c}$, and the marginal distributions $F_j^{\rm obs}$, $F_j$ are known. Here, $R^{\rm c}$ is the correlation matrix between the components of $\mathbf{Z}^{\rm obs}$ given $\mathbf{Z}$, while $R$ (as before) is the correlation matrix between components of $\mathbf{Z}$. $F_j$ is fully determined by the parameter set $\theta_j$, while $F_j^{\rm obs}$ can be computed given $\theta^{\rm c}_j$ and $\theta_j$ via equation (\ref{eq:Fobs}).

The starting point is to write down the conditional probability density of $\mathbf{Z}^{\rm obs}$ given $\mathbf{Z}$. Since $(\mathbf{Z}^{\rm obs}, \mathbf{Z})$ is multivariate normal with correlation matrix $P$ given by (\ref{eq:Pmatrix}), the conditional probability is also multivariate normal \citep{Petersen2012}
\begin{equation}
\label{eq:CondProb}
p(\mathbf{z}^{\rm obs}\vert\mathbf{z}) = \phi_{S - TR^{-1}T^\top}(\mathbf{z}^{\rm obs} - TR^{-1} \mathbf{z}).
\end{equation}

In the following I assume that, for all $j$, the probability distribution of $Z_j^{\rm obs}$ is conditionally independent of $Z_{k\neq{}j}$ given $Z_j$. In other words, I assume that the measured value of variable $j$ does not depend on the \emph{true} values of variables $k\neq{}j$ for a fixed, true value of variable $j$. Note that the measured value of variable $j$ can still depend on the \emph{measured} values of variables $k\neq{}j$ even if the true value of variable $j$ is kept fixed. Hence, the above assumption is not as restrictive as it may at first appear and it allows measurement errors to be correlated.

The assumption that $Z_j^{\rm obs}$  is conditionally independent of $Z_{k\neq{}j}$ given $Z_j$ implies that $TR^{-1}$ is a diagonal matrix. Hence, $T_{ij} = T_{ii} R_{ij}$, or in matrix notation, $T = {\rm diag}(T)\,R$, where ${\rm diag}(T)$ is a diagonal matrix with the same diagonal entries as $T$.

The matrix $Q=S - TR^{-1}T^\top$ is the covariance matrix of the probability distribution of $\mathbf{Z}^{\rm obs}$ given $\mathbf{Z}$. $R^{\rm c}$ is, by definition, the corresponding correlation matrix. Given $T = {\rm diag}(T)\,R$,
\[
Q=S - {\rm diag}(T)\,T^\top = S - {\rm diag}(T)\,R\,{\rm diag}(T).
\]
Hence, ${\rm diag}(Q) = 1 - [{\rm diag}(T)]^2$ and the corresponding correlation matrix is
\[
R^{\rm c} =\left(1 - [{\rm diag}(T)]^2\right)^{-1/2} Q \left(1 - [{\rm diag}(T)]^2\right)^{-1/2}.
\]

To summarize, the matrix $S$ is given as
\begin{multline}
\label{eq:Smatrix}
S =  {\rm diag}(T)\,R\,{\rm diag}(T) \\ +  \left(1 - [{\rm diag}(T)]^2\right)^{1/2}R^{\rm c} \left(1 - [{\rm diag}(T)]^2\right)^{1/2},
\end{multline}
or in index notation
\begin{equation}
\label{eq:Smatrix2}
S_{ij} = R_{ij}T_{ii}T_{jj} + R^{\rm c}_{ij} \sqrt{(1-T^2_{ii})(1-T^2_{jj})}.
\end{equation}

It is instructive to check whether the predictions of equations (\ref{eq:Smatrix}) and (\ref{eq:Smatrix2}) coincide with expectations for a few special cases. First, the equations show that $S=R^{\rm c}$ if the cross-correlations between $\mathbf{Z}^{\rm obs}$ and $\mathbf{Z}$ vanish (i.e., ${\rm diag}(T)=0$). In other words, if measured data are uncorrelated with their true values, then correlations between measured data are solely determined by $R^{\rm c}$ as expected (i.e., by the correlation between measured data given $\mathbf{Z}$). If $\mathbf{Z}^{\rm obs}$ and $\mathbf{Z}$ are maximally correlated (i.e., ${\rm diag}(T)=1$) or anti-correlated (i.e., ${\rm diag}(T)=-1$), then the equations predict $S=R$. Again, this is what one would expect as in this case the measurement uncertainty is zero and the correlation between observables is set by the correlation between $\mathbf{Z}$. Finally, if both $R_{ij}$ and $R_{ij}^{\rm c}$ are equal to one (or both are equal to minus one), then $S_{ij}$ also equals one (or minus one), independent of $T_{ii}$ and $T_{jj}$. In other words, if two variables are perfectly correlated (anti-correlated) and have perfectly correlated (anti-correlated) errors, then their observed values will be perfectly correlated (anti-correlated) independent of the measurement error of either variable.

It is easy to see that $S$ is a correlation matrix provided $R$ and $R^{\rm c}$ are correlation matrices. First, $S$ is symmetric and real (given $-1\leq{}T_{ii}\leq{}1$ for all $i$). Second, its diagonal entries are evidently 1. Finally, $x^\top{}Sx\geq{}0$ for each $K$ vector $x$ (i.e., $S$ is positive semi-definite) because the same is true for each of the two summands on the right hand side. Given that $R$ and $Q$ are positive semi-definite matrices, it also follows that $P$ (which has the shape given by equation (\ref{eq:Pmatrix})) is positive semi-definite given that its upper left block matrix has the form $S = Q + TR^{-1}T^\top$ \citep{Albert1969, Bekker1988}. Furthermore, because $R$ and $S$ are correlation matrices, $P$ is symmetric and has diagonal elements of one. In summary, $S$ and $P$ are correlation matrices provided the input parameters $R$ and $R^{\rm c}$ are correlation matrices.

Measurement errors may be uncorrelated. This case amounts to equating $R^{\rm c}$ with the identity matrix, i.e., $S_{ij} = R_{ij}T_{ii}T_{jj}$ for off-diagonal elements and $S_{ii}=1$ for elements on the diagonal.

Calculating the matrix elements of $S$ via equations (\ref{eq:Smatrix}) or (\ref{eq:Smatrix2}) requires knowledge of the diagonal elements of $T$ in addition to $R$ and $R^{\rm c}$. These diagonal elements can be derived as follows. Equations  (\ref{eq:Pyobsy})-(\ref{eq:Pmatrix}) are completely analogous for any subset of variables from $\{Y_1^{\rm obs},\ldots,Y_K^{\rm obs},Y_1,\ldots,Y_K\}$, i.e., after marginalizing over all the other variables. Hence, 
\[
p(y^{\rm obs}_j, y_j) = \phi_{\tilde{T}_j}(z^{\rm obs}_j, z_j) \frac{f^{\rm obs}_j(y^{\rm obs}_j)}{\phi(z^{\rm obs}_j)}\frac{f_j(y_j)}{\phi(z_j)}
\]
with the $2\times{}2$ sub-matrix of $P$:
\[
\tilde{T}_j = \begin{pmatrix}1 & T_{jj} \\ T_{jj} & 1\end{pmatrix}
\]
and
\[
p(y_j) = \phi(z_j)\frac{f_j(y_j)}{\phi(z_j)}=f_j(y_j).
\]

Hence, the probability density of $y^{\rm obs}_j$ conditional on $y_j$ is
\begin{equation}
\label{eq:Tdiag}
\begin{split}
p(y^{\rm obs}_j \vert y_j) &= \frac{p(y^{\rm obs}_j, y_j)}{p(y_j)} = \frac{\phi_{\tilde{T}_j}(z^{\rm obs}_j, z_j) }{\phi(z_j)}\frac{f^{\rm obs}_j(y^{\rm obs}_j)}{\phi(z^{\rm obs}_j)} \\
&=  \phi_{1-T^2_{jj}}(z^{\rm obs}_j - T_{jj}\,z_j) \frac{f^{\rm obs}_j(y^{\rm obs}_j)}{\phi(z^{\rm obs}_j)}.
\end{split}
\end{equation}
The last equality follows from equation (\ref{eq:CondProb}).

Equation (\ref{eq:Tdiag}) holds for any given $z_j$ and $z^{\rm obs}_j$ (and the corresponding $y_j$ and $y^{\rm obs}_j$). For instance, choosing $z_j=z^{\rm obs}_j=0$ results in
\begin{equation}
\frac{1}{1 - T^2_{jj}} = \frac{p({\rm med}(Y^{\rm obs}_j)|{\rm med}(Y_j))}{ f^{\rm obs}_j({\rm med}(Y^{\rm obs}_j))},
\end{equation}
where ${\rm med}(X)$ is the median value of random variable $X$. This ansatz has two disadvantages. First, it requires knowledge of the median of $Y^{\rm obs}_j$. In general, this can be computationally costly to calculate. Secondly, and more importantly, this equation has two solutions. It is thus only useful if the sign of $T_{jj}$ is known.

A better approach is to evaluate $p(y^{\rm obs}_j \vert y_j)$ for two different values of $y_j$ ($y^+_j$ and $y^-_j$) and to compute the logarithm of the ratio:
\begin{equation}
\label{eq:Tjj}
\begin{split}
a_j\equiv{}\ln{}\frac{p(y^{\rm obs}_j \vert y^+_j)}{p(y^{\rm obs}_j \vert y^-_j)} &=\ln{}\frac{ \phi_{1-T^2_{jj}}(z^{\rm obs}_j - T_{jj}\,z^+_j)}{ \phi_{1-T^2_{jj}}(z^{\rm obs}_j - T_{jj}\,z^-_j)}  \\
&= - \frac{(z^{\rm obs}_j - T_{jj}\,z^+_j)^2 - (z^{\rm obs}_j - T_{jj}\,z^-_j)^2}{2(1 - T^2_{jj})} \\
&= \Delta{}z_j \, \frac{T_{jj}z^{\rm obs}_j - T^2_{jj}\,\bar{z}_j}{1 - T^2_{jj}}.
\end{split}
\end{equation}
Here, $\Delta{}z_j = (z^+_j-z^-_j)$ and $\bar{z}_j=(z^+_j+z^-_j)/2$. 

If $a_j = \bar{z}_j\Delta{}z_j$, then $T_{jj}=\bar{z}_j/z^{\rm obs}_j$. In general, the quadratic equation given by (\ref{eq:Tjj}) can have zero, one, or two real solutions for $T_{jj}$. Fortunately, this equation has exactly one real solution with $-1\leq{}T_{jj}\leq{}1$:
\begin{equation}
\label{eq:Tjj2}
T_{jj} = - b_j \pm{} \sqrt{b_j^2 + 1},\textrm{ with }b_j=\frac{z^{\rm obs}_j \Delta{}z_j}{2a_j},
\end{equation}
provided $\bar{z}=0$ and\footnote{The case $b_j=0$ corresponds to $\Delta{}z_j=0$ or $z^{\rm obs}_j=0$. In this case a different value should be chosen for $y^+_j$ and $y^{\rm obs}_j$, respectively.}  $b_j\neq{}0$. The $+$ sign ($-$ sign) applies if $b_j\geq{}0$ ($b_j\leq{}0$).

In practice, a reasonable strategy is to pick a value for $y^+_j$ such that $p(y^{\rm obs}_j \vert y^+_j)>0$ and then to compute $z^+_j = \Phi^{-1}(F_j(y^+_j))$, $z^-_j = -z^+_j$ (thus ensuring $\Delta{}z_j=0$), $y^-_j=F_j^{-1}(\Phi(z^-_j))$, 
$z^{\rm obs}_j=\Phi^{-1}(F^{\rm obs}_j(y^{\rm obs}_j))$, and finally $p(y^{\rm obs}_j \vert y^-_j)$. The natural choice $y^+_j=y^{\rm obs}_j$ results in $b_j = z_j^{\rm obs}z^+_j / a_j$. Note that, in general, $z^{\rm obs}_j\neq{}z^+_j$ even if $y^{\rm obs}_j=y^+_j$. Equation (\ref{eq:Tjj2}) provides $T_{jj}$ for each $j$, $S$ is computed via equation (\ref{eq:Smatrix}), and $p(\mathbf{y}^{\rm obs})$ is given by equation (\ref{eq:pyobs2}).

\subsection{Censored and missing data}
\label{sect:CensoredAndMissingData}
The approach outlined in the previous section to compute $p(\mathbf{y}^{\rm obs})$ can be easily modified to account for data censoring and for a certain class of missing data. 

Many data sets contain missing data entries. The probability that a given data entry is missing may depend both on the values of the observed data and on the values of missing or censored data. In this fully general scenario, treating missing data requires an explicit `missing data' model. However, in many situations data values are `missing at random' (MAR). In MAR the probability that a given observation is missing may depend on other measured data (but it may not depend on unmeasured data). Data that is MAR is a much more general case than data that is `missing completely at random' (MCAR). Fortunately, under some additional, weaker conditions, data that is MAR can be treated properly without a missing data model, i.e., the missing data model is ignorable. The approach described below assumes that data is MAR and that the missing data model is ignorable.

A data value is censored if its exact value is unknown but it is known to fall into a given interval. Data censoring arises in a variety of contexts and scientific fields, such as due to drop outs in biomedical research and epidemiology \citep{Collett2003, Ahrens2005}, and via detection limits in astronomical surveys \citep{Feigelson2012}. Data censoring plays a paramount role in survival analysis \citep{Kalbfleisch2002, CRC2019}.

Data censoring and missing data are related concepts. However, whether a data point is censored generally depends on the data value itself. Hence, censoring amounts to a `missing not at random' (MNAR) scenario and an explicit model for censoring is needed. The censoring model adopted in this paper is based on the concept of upper and lower detection limits. Specifically, a measurement that returns a value below or above a given detection limit is flagged as censored and the measured value (if provided) is discarded.  Mathematically, if the probability density of a particular observable $y$ is $p(y)$, then the probability that the value of that particular observable is below the detection limit $y^{\rm lo}$ is $\int_{-\infty}^{y^{\rm lo}} dy p(y)$. The probability of $y$ being above a detection limit $y^{\rm up}$ is similarly modeled as $\int_{y^{\rm up}}^\infty dy p(y)$. This basic approach for computing probabilities in the presence of censoring is formalized below.

In what follows, $\theta$ parametrizes the generative model of producing observational data $\mathbf{y}^{\rm obs}$, $\phi$ parametrizes the missing data model, and $\mathbf{\Omega}^{\rm lim}$ is a set of intervals of the form $(-\infty,y^{\rm lo}_j)$ or $(y^{\rm up}_j, \infty)$ indicating lower or upper detection limits.
The observational data set can be split into data with measured values ($\mathbf{y}^{\rm meas}$), missing values ($\mathbf{y}^{\rm miss}$), and censored but non-missing values ($\mathbf{y}^{\rm cen}$). A binary indicator variable ($\mathbf{m}$) encodes whether a data value is missing ($m_{j}=1$ if $y^{\rm obs}_{j}$ is missing, otherwise $m_{j}=0$). An analogous binary variable ($\mathbf{c}$) is used to indicate censoring.

In the previous sections I derived the probability (density) of measuring $\mathbf{y}^{\rm obs}$ given model parameters $\theta$ for the special case $\mathbf{y}^{\rm obs}=\mathbf{y}^{\rm meas}$. In the presence of missing and censored data, the relevant quantity of interest becomes $p(\mathbf{y}^{\rm meas}, \mathbf{m}, \mathbf{c}\, \vert{}\, \theta, \mathbf{\Omega}^{\rm lim})$.

The starting point is to marginalize the probability density
\begin{multline}
\label{eq:cenmisful}
p(\mathbf{y}^{\rm meas}, \mathbf{y}^{\rm cen}, \mathbf{y}^{\rm miss}, \mathbf{m}, \mathbf{c}\, \vert{}\, \theta, \phi, \mathbf{\Omega}^{\rm lim}) = \\
\begin{split}
&p(\mathbf{m}\,\vert\, \mathbf{c},\mathbf{y}^{\rm meas}, \mathbf{y}^{\rm cen}, \mathbf{y}^{\rm miss},\theta, \phi, \mathbf{\Omega}^{\rm lim}) \\
&\times{} p(\mathbf{c}, \mathbf{y}^{\rm meas}, \mathbf{y}^{\rm cen}, \mathbf{y}^{\rm miss}\,\vert\,\theta, \phi, \mathbf{\Omega}^{\rm lim}) \\
\end{split}
\end{multline}
over $\mathbf{y}^{\rm cen}$ and $\mathbf{y}^{\rm miss}$.
The first term on the right hand side (the missing data model) can be substantially simplified if data is missing at random (MAR), i.e., if $\mathbf{m}$ only depends on known data ($\mathbf{y}^{\rm meas}, \mathbf{c}$). Furthermore, I assume that $\mathbf{m}$ is independent of the generative model parameters ($\theta$). Under these simplifying assumptions, and after marginalizing, I obtain:
\begin{multline}
\label{eq:cenmismarg1}
p(\mathbf{y}^{\rm meas}, \mathbf{m}, \mathbf{c}\, \vert{}\, \theta, \phi, \mathbf{\Omega}^{\rm lim}) = \\
p(\mathbf{m}\,\vert\, \mathbf{c},\mathbf{y}^{\rm meas}, \phi, \mathbf{\Omega}^{\rm lim}) 
\times{} p(\mathbf{c}, \mathbf{y}^{\rm meas}\,\vert\,\theta, \phi, \mathbf{\Omega}^{\rm lim}) 
\end{multline}

A direct way to calculate $p(\mathbf{c}, \mathbf{y}^{\rm meas}\,\vert\,\theta, \phi, \mathbf{\Omega}^{\rm lim})$ is to marginalize over $\mathbf{y}^{\rm cen}$ and $\mathbf{y}^{\rm miss}$ in the following term
\begin{multline}
\label{eq:cenmismarg3}
p(\mathbf{c}, \mathbf{y}^{\rm meas}, \mathbf{y}^{\rm cen},  \mathbf{y}^{\rm miss} \,\vert\,\theta, \phi, \mathbf{\Omega}^{\rm lim}) =\\
\begin{split}
&p(\mathbf{c} \,\vert\, \mathbf{y}^{\rm meas}, \mathbf{y}^{\rm cen}, \mathbf{y}^{\rm miss}, \theta, \phi, \mathbf{\Omega}^{\rm lim}) \times{} \\
&p(\mathbf{y}^{\rm meas}, \mathbf{y}^{\rm cen}, \mathbf{y}^{\rm miss} \,\vert\, \theta, \phi, \mathbf{\Omega}^{\rm lim}).
\end{split}
\end{multline}

I assume the following model for data censoring $p(\mathbf{c} \,\vert\, \mathbf{y}^{\rm meas}, \mathbf{y}^{\rm cen}, \mathbf{y}^{\rm miss}, \theta, \phi, \mathbf{\Omega}^{\rm lim}) = \prod_{j}h_j$, where $h_j=H(y^{\rm lo}_{j} - y_{j}^{\rm cen})$ if a lower detection limit is provided (for variables $\mathbf{y}^{\rm cen,lo}$) and $h_j=H(y_{j}^{\rm cen} - y^{\rm up}_{j})$ if an upper detection limit is given (for variables $\mathbf{y}^{\rm cen,lo}$), with  $H$ the Heaviside step function. In other words, the probability that values are censored is equal to one if these values are outside their detection limits, and zero otherwise. To simplify the treatment of detection limits, any given variable may have lower and/or upper limits, but not both for the same data point, i.e., for any observation $\mathbf{y}^{\rm cen}=(\mathbf{y}^{\rm cen,lo}, \mathbf{y}^{\rm cen,up})$. Furthermore, I assume that the probability density of $\mathbf{y}^{\rm obs}$ is independent of the parameters of the missing and censoring models.

With these assumptions equation (\ref{eq:cenmismarg3}) simplifies to
\begin{multline}
\label{eq:cenmismarg2}
p(\mathbf{c}, \mathbf{y}^{\rm meas}\,\vert\,\theta, \phi, \mathbf{\Omega}^{\rm lim}) = p(\mathbf{c}, \mathbf{y}^{\rm meas}\,\vert\,\theta, \mathbf{\Omega}^{\rm lim})  \\
\begin{split}
&= \int_{\mathbf{\Omega}^{\rm lim}} d\mathbf{y}^{\rm cen} \int_{-\infty}^{\infty} d\mathbf{y}^{\rm miss}  p(\mathbf{y}^{\rm meas}, \mathbf{y}^{\rm cen}, \mathbf{y}^{\rm miss} \,\vert\, \theta) \\
&= \int_{-\infty}^{\mathbf{y}^{\rm lo}} d\mathbf{y}^{\rm cen, lo} \int_{\mathbf{y}^{\rm up}}^\infty d\mathbf{y}^{\rm cen, up}\,   p(\mathbf{y}^{\rm meas}, \mathbf{y}^{\rm cen} \,\vert\, \theta). \\
\end{split}
\end{multline}

$p(\mathbf{y}^{\rm meas}, \mathbf{m}, \mathbf{c}\, \vert{}\, \theta, \mathbf{\Omega}^{\rm lim})$ can be computed from equation (\ref{eq:cenmismarg1}) by multiplying its left hand side with a prior for $\phi$ and marginalizing over $\phi$. The result is
\begin{equation}
p(\mathbf{y}^{\rm meas}, \mathbf{m}, \mathbf{c}\, \vert{}\, \theta, \mathbf{\Omega}^{\rm lim}) = p(\mathbf{m}\,\vert\, \mathbf{c},\mathbf{y}^{\rm meas}, \mathbf{\Omega}^{\rm lim})\times{}p(\mathbf{c}, \mathbf{y}^{\rm meas}\,\vert\,\theta, \mathbf{\Omega}^{\rm lim}).
\end{equation}
The first term on the right hand side is the missing data model, while the second term on the right hand side is given by equation (\ref{eq:cenmismarg2}). With the assumptions that were made (primarily that data is MAR), the missing data model is `ignorable',
i.e., it only amounts to an irrelevant constant of proportionality for the likelihood of $\theta$. Furthermore, it has no bearing on the posterior distribution of the model parameters $\theta$. This can be seen as follows
\begin{multline}
p(\theta \,\vert\, \mathbf{y}^{\rm meas}, \mathbf{m}, \mathbf{c}, \mathbf{\Omega}^{\rm lim}) \\
\begin{split}
&= \frac{p(\mathbf{y}^{\rm meas}, \mathbf{m}, \mathbf{c}\, \vert{}\, \theta, \mathbf{\Omega}^{\rm lim})p(\theta\,\vert\,\mathbf{\Omega}^{\rm lim})}{p(\mathbf{y}^{\rm meas}, \mathbf{m}, \mathbf{c} \,\vert\, \mathbf{\Omega}^{\rm lim})} \\
&= \frac{ p(\mathbf{m}\,\vert\, \mathbf{c},\mathbf{y}^{\rm meas}, \mathbf{\Omega}^{\rm lim}) }{p(\mathbf{y}^{\rm meas}, \mathbf{m}, \mathbf{c} \,\vert\, \mathbf{\Omega}^{\rm lim})} p(\mathbf{c}, \mathbf{y}^{\rm meas}\,\vert\,\theta, \mathbf{\Omega}^{\rm lim}) p(\theta\,\vert\,\mathbf{\Omega}^{\rm lim}) \\
&= \frac{ p(\mathbf{c}, \mathbf{y}^{\rm meas}\,\vert\,\theta, \mathbf{\Omega}^{\rm lim}) p(\theta\,\vert\,\mathbf{\Omega}^{\rm lim})}{p(\mathbf{y}^{\rm meas}, \mathbf{c} \,\vert\, \mathbf{\Omega}^{\rm lim})}\\
&=p(\theta \,\vert\, \mathbf{y}^{\rm meas}, \mathbf{c}, \mathbf{\Omega}^{\rm lim}).
\end{split}
\end{multline}
Hence, $p(\mathbf{c}, \mathbf{y}^{\rm meas}\,\vert\,\theta, \mathbf{\Omega}^{\rm lim})$ can be used as the likelihood function of $\theta$ for a given observation in the presence of missing and censored data values.

Applying equation (\ref{eq:cenmismarg2}) to equation (\ref{eq:pyobs2}) leads to
\begin{multline}
\label{eq:likelihood}
p(\mathbf{c}, \mathbf{y}^{\rm meas}\,\vert\,\theta, \mathbf{\Omega}^{\rm lim})   \\
\begin{split}
= & \int_{-\infty}^{\mathbf{y}^{\rm lo}} d\mathbf{y}^{\rm cen,lo}  \int_{\mathbf{y}^{\rm up}}^\infty  d\mathbf{y}^{\rm cen,up} \int_{-\infty}^{\infty} d\mathbf{y}^{\rm miss}  
\phi_S(\mathbf{z}^{\rm obs})  \prod_{j=1}^K \frac{f^{\rm obs}_j(y^{\rm obs}_j)}{\phi(z^{\rm obs}_j)} \\
= & \prod_{j=1}^{K^{\rm meas}} \frac{f^{\rm obs}_j(y^{\rm meas}_{j})}{\phi(z^{\rm meas}_{j})} \\
& \times \int_{-\infty}^{\mathbf{z}^{\rm lo}}  d\mathbf{z}^{\rm cen,lo}   \int_{\mathbf{z}^{\rm up}}^\infty  d\mathbf{z}^{\rm cen, up} \int_{-\infty}^{\infty} d\mathbf{z}^{\rm miss} \phi_S(\mathbf{z}^{\rm meas},  \mathbf{z}^{\rm cen}, \mathbf{z}^{\rm miss}).
\end{split}
\end{multline}
Here, I assume without loss of generality that the $K$ $z^{\rm obs}_1,\ldots,z^{\rm obs}_K$ variables are ordered such that $\mathbf{z}^{\rm meas}$ are the first $K^{\rm meas}$ variables, $\mathbf{z}^{\rm cen}$ are the next $K^{\rm cen}$ variables, and $\mathbf{z}^{\rm miss}$ are the last $K^{\rm miss}$ variables with $K = K^{\rm meas}+K^{\rm cen}+K^{\rm miss}$.

It is straightforward to compute the incomplete multidimensional integral of the multivariate normal distributions in equation (\ref{eq:likelihood}) using the decomposition of the joint distribution into a product of marginal and conditional distributions, see equation (\ref{eq:CondProb}). I start by rewriting the $K\times{}K$ matrix $S$ in block-matrix form as
\[
S = \begin{pmatrix} S_1 & T_1 \\ T_1^\top & U_1\end{pmatrix}, \textrm{ with } S_1 = \begin{pmatrix}S_2 & T_2\\T_2^\top & U_2\end{pmatrix},
\]
where $U_1, U_2, T_2$, and $S_2$ are $K^{\rm miss}\times{}K^{\rm miss}$, $K^{\rm cen}\times{}K^{\rm cen}$, $K^{\rm meas}\times{}K^{\rm cen}$, and $K^{\rm meas}\times{}K^{\rm meas}$ matrices, respectively. All these matrices are known given that $S$ can be calculated as outlined in section \ref{sect:S}.

It then follows that 
\begin{multline}
\label{eq:multidimintegral}
\int_{-\infty}^{\mathbf{z}^{\rm lo}}  d\mathbf{z}^{\rm cen,lo} \int_{\mathbf{z}^{\rm up}}^\infty  d\mathbf{z}^{\rm cen,up} \int_{-\infty}^{\infty} d\mathbf{z}^{\rm miss} \phi_S(\mathbf{z}^{\rm meas}, \mathbf{z}^{\rm miss}, \mathbf{z}^{\rm cen}) \\
\begin{split}
=&\int_{-\infty}^{\mathbf{z}^{\rm lo}}  d\mathbf{z}^{\rm cen,lo} \int_{\mathbf{z}^{\rm up}}^\infty  d\mathbf{z}^{\rm cen,up}\, \phi_{S_1}(\mathbf{z}^{\rm meas}, \mathbf{z}^{\rm cen}) \\
=&\,\phi_{S_2}(\mathbf{z}^{\rm meas}) \int_{-\infty}^{\mathbf{z}^{\rm lo}}  d\mathbf{z}^{\rm cen,lo} \int_{\mathbf{z}^{\rm up}}^\infty  d\mathbf{z}^{\rm cen,up} \\
&\times \, \phi_{U_2 - T_2^\top{}S_2^{-1}T_2^{\phantom{\top}}}(\mathbf{z}^{\rm cen} - T_2^\top{} S_2^{-1}\mathbf{z}^{\rm meas}) \\
=&\,\phi_{S_2}(\mathbf{z}^{\rm meas}) \int_{-\infty}^{\mathbf{\tilde{z}}^{\rm lo}}  d\mathbf{\tilde{z}}^{\rm cen}
 \phi_{U_2 - T_2^\top{}S_2^{-1}T_2^{\phantom{\top}}}(\mathbf{z}^{\rm cen} - T_2^\top{} S_2^{-1}\mathbf{z}^{\rm meas}) \\
=&\,\phi_{S_2}(\mathbf{z}^{\rm meas}) \int_{-\infty}^{\mathbf{\tilde{z}}^{\rm lo}}  d\mathbf{\tilde{z}}^{\rm cen}
 \phi_{A(U_2 - T_2^\top{}S_2^{-1}T_2^{\phantom{\top}})\,A}(\mathbf{\tilde{z}}^{\rm cen} - AT_2^\top{} S_2^{-1}\mathbf{z}^{\rm meas}) \\
 =&\, \phi_{S_2}(\mathbf{z}^{\rm meas})\times{}\Phi_{A(U_2 - T_2^\top{}S_2^{-1}T_2^{\phantom{\top}})\,A}(\mathbf{\tilde{z}}^{\rm lo} - AT_2^\top{} S_2^{-1}\mathbf{z}^{\rm meas}),
\end{split}
\end{multline}
with $\mathbf{\tilde{z}}^{\rm lo}=(\mathbf{z}^{\rm lo}, -\mathbf{z}^{\rm up})$ and $\mathbf{\tilde{z}}^{\rm cen}=(\mathbf{z}^{\rm cen,lo}, -\mathbf{z}^{\rm cen,up})=A\mathbf{z}^{\rm cen}$, i.e., $A$ is a diagonal matrix with eigenvalues $1$ and $-1$.

The likelihood $\mathcal{L}_i(\theta)$ of $\theta$ for a given observation $i$ (up to an irrelevant multiplicative constant) can be calculated from equations (\ref{eq:likelihood}) and (\ref{eq:multidimintegral}).

As individual observations are assumed to be independent from each other, the full likelihood of $\theta$ for a set of $N$ observations is
\begin{multline}
\label{eq:fullLikelihood}
\mathcal{L}(\theta) = \prod_{i=1}^{N}\mathcal{L}_i(\theta) = \prod_{i=1}^{N} p(\mathbf{c}_i, \mathbf{y}^{\rm meas}_i\,\vert\,\theta, \mathbf{\Omega}^{\rm lim}_i) = \\
\begin{split}
 &\prod_{i=1}^{N}\phi_{S_2}(\mathbf{z}^{\rm meas}_i) \times{}\Phi_{A(U_2 - T_2^\top{}S_2^{-1}T_2^{\phantom{\top}})\,A}(\mathbf{\tilde{z}}^{\rm lo}_i - AT_2^\top{} S_2^{-1}\mathbf{z}^{\rm meas}_i) \\
& \quad{}\quad{}\times{} \prod_{j=1}^{K^{\rm meas}_i} \frac{f^{\rm obs}_{ij}(y^{\rm meas}_{ij})}{\phi(z^{\rm meas}_{ij})}
\end{split}
\end{multline}
Most of the terms and variables in the equation above depend on the given observation indexed by $i$. For instance, whether a given data variable $j$ is censored is observation dependent as are the matrix $A$ and detection limits $\mathbf{\tilde{z}}^{\rm lo}$. Furthermore, the $S$ matrix depends on the error model (i.e., on $R^{\rm c}$ and $f^{\rm obs}$) and the error model may vary with observation. Hence, in the general case, $S, U_2, T_2$, and $S_2$ have to be calculated for each observation.
 
 \subsection{Special case: Multivariate normal distribution with normally distributed errors.} 
\label{sect:MultivariateNormal}

An important special case of the approach outlined in the previous sections arises if the true data $\mathbf{Y}$ has a multivariate normal distribution and the observational errors are also normally distributed with zero mean. In this case the likelihood for a single observation is a multivariate normal distribution given that
\[
p(\mathbf{y}^{\rm obs}) = \int p(\mathbf{y}^{\rm obs}\vert\mathbf{y}) p(\mathbf{y}) d\mathbf{y}
\]
is simply a convolution of two multivariate normal distributions. Specifically, $p(\mathbf{y}^{\rm obs}) = \phi_W(\mathbf{y}^{\rm obs} - \bm{\mu})$, where $\bm{\mu}$ is the expected value of $\mathbf{Y}$ and $W=\Sigma + \Sigma^{\rm c}$ is the sum of the covariances of the two multivariate normal distributions. I show in the appendix that the approach outlined in the previous sections reproduces this basic result.
 
\subsection{Modeling outliers}
\label{sect:Outliers}

Model fitting can be sensitive to even a small number of outliers, i.e., data points that are unlikely to be generated by the assumed model. Outliers frequently arise as a consequence of model incompleteness. Here, outlying data points might belong to a small but influential, data component that had been neglected in the generative model. A natural way to account for such outliers is to enrich the adopted generative model to explicitly model the outlier distribution (e.g., \citealt{Hogg2010}).

To be more specific, assume that ``good'' data and outliers are distributed with distributions $p_{\rm good}$ and $p_{\rm out}$, each with their own set of parameters ($\theta$ and $\phi$). Also, let $\psi$ be the probability that any given data point $\mathbf{y}$ is an outlier. The probability density of $\mathbf{y}$ is then described by a mixture model
\[
p_{\rm mix}(\mathbf{y}\vert{}\theta,\phi,\psi) = (1-\psi) p_{\rm good}(\mathbf{y}\vert{}\theta) + \psi p_{\rm out}(\mathbf{y}\vert{}\phi).
\]
Hence, outliers of this kind are properly accounted for by replacing the original probability density with $p_{\rm mix}$ and by adding the parameters $\phi$, $\psi$.

\subsection{Regression problems} 
\label{sect:Regression}

Regression problems such as those introduced in section~\ref{sect:Motivation} can be easily studied with the help of latent variables. Let $p(\mathbf{y^{\rm obs}}\vert{}\theta)$ be the probability density of observing values $\mathbf{y^{\rm obs}}$ if the true model has parameters $\theta$ and $R$. The model parameters in question refer to the location, scale, and shape parameters of the 1-dimensional marginal distributions of $\{y_j\}_{j=1,\ldots{},K}$ and their correlations. Let further $p(\mathbf{y^{\rm obs}}\vert{}t,\omega, R)=p(\mathbf{y^{\rm obs}}\vert{}\theta(t,\omega), R)$ be the probability density after re-parametrizing $\theta$ as a function of a latent (scalar or vector) variable $t$ and another set of parameters $\omega$. Here, $\omega$ may refer to, e.g., the slope and intercept of a linear regression line (see section~\ref{sect:Motivation}). The likelihood of $(\omega, R)$ is simply:
\begin{equation}
\label{eq:regression}
\begin{split}
p(\mathbf{y^{\rm obs}}\vert{}\omega, R)&=\int dt p(t\vert{}\omega, R) p(\mathbf{y^{\rm obs}}\vert{}t, \omega, R)\\
&=\frac{1}{p(\omega, R)}\int dt p(\omega, R\vert{}t)p(t) p(\mathbf{y^{\rm obs}}\vert{}\theta(t,\omega), R),
\end{split}
\end{equation}
and, in case $p(\omega, R\vert{}t)$ does not depend on $t$,
\begin{equation}
\label{eq:regression2}
p(\mathbf{y^{\rm obs}}\vert{}\omega, R) = \int dt p(t) p(\mathbf{y^{\rm obs}}\vert{}\theta(t, \omega), R).
\end{equation}
The expression  $p(\mathbf{y^{\rm obs}}\vert{}\theta(t,\omega), R)$ can be evaluated for each  $\theta(t,\omega)$ and $R$ as outlined in the previous sections. In practice, this calculation is done by \leopy{}.
Subsequently, equations~(\ref{eq:regression}) or (\ref{eq:regression2}) may be solved by direct numerical integration if the dimensionality of the latent variable $t$ is small, e.g., if $t$ is a scalar. Otherwise, alternative approaches that are more adequate in higher-dimensional spaces, such as Monte-Carlo integration, should be used. 

\subsection{Caveats}

The presented approach in its present form has a number of limitations, some of which are already mentioned in the beginning of section \ref{sect:Method}.  These include the assumption that the marginal distributions are continuous probability distributions and that the correlation structure is well described by a Gaussian copula. Clearly, for some problems either assumption may not be justified.

Other limitations include the treatment of missing data. In its present form, only data `missing at random` can be correctly accounted for. Relaxing this assumption would require the user to provide a specific model for missing data. 

Finally, there is the matter of computing speed. Likelihood calculations can be computationally intensive depending on the problem at hand. For instance, a single likelihood calculation for the data set shown in the left panel of Fig.~\ref{fig:example_regression} requires about 30 ms on the author's laptop computer, while a likelihood calculation for the xGASS data set, see section \ref{sect:DataSet}, requires about 160 ms. Given that a likelihood maximization requires of hundreds of likelihood calls, finding the most likely parameters can take minutes or hours on a single core. Fortunately, the computation can be trivially parallelized given that the likelihood calculation of different observations can be performed independently from each other. Hence, if sufficient computing power is available, the wall-clock time can be reduced significantly.

\section{\leopy{}: Usage}
\label{sect:Code}

The mathematical framework developed in section \ref{sect:Method} is fully implemented in the Python package \leopy. The package is available from the Python Package Index and it can be installed via
\begin{lstlisting}[language=bash]
	pip install leopy-stat
\end{lstlisting}
from the command line. Alternatively, it can be cloned from the GitHub repository hosted at \url{https://github.com/rfeldmann/leopy} and installed via
\begin{lstlisting}[language=bash]
	python setup.py install
\end{lstlisting}
from the package directory. \leopy{} requires the SciPy, NumPy, and pandas packages \citep{Jones2001, Oliphant2006, mckinney-proc-scipy-2010, VanderWalt2011} which are automatically installed if necessary.

The source distribution contains a large number of examples and test cases, including those presented in section \ref{sect:Motivation}. The test suite can be run from within the package directory by typing 
\begin{lstlisting}[language=bash]
	python setup.py test
\end{lstlisting}
in the command line.

In the sections below I highlight a few common use cases. Further information can be found in the examples distributed with the code, the in-code documentation, and the HTML software documentation created with the Sphinx documentation generator. 

\subsection{A minimal example}
\label{sect:minimal example}

Using \leopy{} in a Python program is very simple. A script may look like the below:
\begin{lstlisting}[language=Python]
import leopy

d = {'v0': [1, 2], 'e_v0': [0.1, 0.2],
     'v1': [3, 4], 'e_v1': [0.1, 0.1]}
obs = leopy.Observation(d, 'testdata')
    
like = leopy.Likelihood(obs, 
    p_true='gamma', p_cond='norm')
    
like.p([0.5, 0.7], [1, 2], 
       shape_true=[1.4, 2])
\end{lstlisting}

The \leopy{} library and other required packages are loaded in a first step. Subsequently, an observational data set is created containing the observed values of two variables and the associated measurement uncertainties. The observational data can be provided as a dictionary (see above) or in form of a pandas table. The observational data set is then stored in an instance of the container class `Observation'. During this step various consistency checks are run on the data set. 

Subsequently, an instance of the class `Likelihood' is created. The parameter families of the 1-dimensional distributions ($f_j$ and $f_j^{\rm c}$, see section \ref{sect:Method}) are specified at this time. In the example above, the marginal distributions of both latent (true) data variables are assumed to follow a gamma distribution, while measurement errors are assumed to have a (bivariate) normal distribution.

Finally, the function p() is called with the parameters of the true data distributions as input. The first parameter is a list of location parameters (one for each variable) of the marginal distributions of the true variables. The second parameter is a corresponding list of scale parameters. Shape parameters of the gamma distributions are provided with the \verb|shape_true| option. 

The output of this function is the probability density of the observational data given the model parameters for each observation. This output can be used, e.g., to find the most likely parameters with standard minimization algorithms such as the ones provided by scipy.optimize.minimize.

\subsection{Correlated measurement errors}

In the example above, the measurement errors of the two variables are uncorrelated. A correlation can be easily specified, however, by adding entries of the form \verb|r_<variable0>_<variable1>| to the data set. For instance, if I modify the following line in the example above
\begin{lstlisting}[language=Python]
d = {'v0': [1, 2], 'e_v0': [0.1, 0.2],
     'v1': [3, 4], 'e_v1': [0.1, 0.1],
     'r_v0_v1': [0.5, -0.8]}
\end{lstlisting}
then correlation matrix of the measurement errors has off-diagonal entries of 0.5 (-0.8) for the first (second) observation. Strictly speaking, the correlation matrix specifies the correlation between the components of the \emph{normalized} observed variables $\mathbf{Z}^{\rm obs}$ given the normalized true data variables $\mathbf{Z}$. However, this correlation matrix is also the correlation matrix between $\mathbf{Y}^{\rm obs}$ given $\mathbf{Y}$ as long as the measurement errors are multivariate normal random variates (\verb|p_cond='norm'|).

\subsection{Correlated data}

\leopy{} models correlations between data variables $\mathbf{Y}$ with Gaussian copulas. The correlation structure is encapsulated in the correlation matrix $R$ that describes the joint normal distribution of the normalized variables $\mathbf{Z}$.
$R$ can be specified when calling the function p(). 

\begin{lstlisting}[language=Python]
like.p([0.5, 0.7], [1, 2], 
       shape_true=[1.4, 2],
       R_true=[[1, 0.2], [0.2, 1]])
\end{lstlisting}

\subsection{Censored and missing data}

Censored values can be specified by adding entries of the form \verb|c_<variable>| to the data set and by providing lower and/or upper limits via \verb|l_<variable>| and/or \verb|u_<variable>|. More specifically, \leopy{} considers an observation of a variable censored if \verb|c_<variable>| is given \emph{and} equals 1 (or True).  For instance, in the data set below
 \begin{lstlisting}[language=Python]
from numpy import nan
d = {'v0': [1, nan], 'e_v0': [0.1, 0.2],
     'v1': [nan, 4], 'e_v1': [nan, 0.1],
     'c_v0': [0, 1], 'l_v0': [nan, 1.3]}
\end{lstlisting}
the second observation of variable `v0' is censored with a lower detection limit of 1.3, while the first observation of variable `v0'  is not censored. \leopy{} ignores the value of \verb|v_<variable>| for censored data and it ignores the values of \verb|l_<variable>| and \verb|u_<variable>| for non-censored data. Measurement uncertainties have to be specified for censored values unless the data is assumed to be free of measurement errors (i.e., \verb|p_cond| is set to \verb|None|).

Values are considered missing if they are non-censored and \verb|v_<variable>| equals `nan'. For instance, in the example above, the first observation of variable `v1' is a missing value, while in the example below
\begin{lstlisting}[language=Python]
from numpy import nan 
d = {'v0': [1, nan], 'e_v0': [0.1, nan],
     'v1': [nan, 4], 'e_v1': [nan, 0.1]}
\end{lstlisting}
both the second observation of `v0' and the first observation of variable `v1' are missing. Measurement uncertainties must be provided for all non-missing data entries unless  \verb|p_cond| is set to \verb|None|.

\subsection{Outliers}
As discussed in section \ref{sect:Outliers}, outliers can be easily accounted for by adding one (or more) additional components to the generative model. For example, the class `gamma\_lognorm` defined in the leopy.stats module adds a lognormal component to a gamma distribution. Replacing a gamma distribution with a gamma--lognormal mixture is thus very straightforward. For example, lognormally distributed outliers in the variable 'v1' can be modeled by changing the minimal example of section \ref{sect:minimal example} into

\begin{lstlisting}[language=Python, basicstyle=\footnotesize\ttfamily]
import leopy

d = {'v0': [1, 2], 'e_v0': [0.1, 0.2],
     'v1': [3, 4], 'e_v1': [0.1, 0.1]}
obs = leopy.Observation(d, 'testdata')
    
like = leopy.Likelihood(obs, 
    p_true=['gamma', leopy.stats.gamma_lognorm],
    p_cond='norm')
    
like.p([0.5, 0.7], [1, 2], 
       shape_true=[1.4, [2, 0.3, 0.9, 0.4]])
\end{lstlisting}

It is very easy to define similar mixed models for other use cases. The reader is referred to the leopy.stats module for specific examples.

\subsection{Parallelization}
\leopy{} supports parallel processing in a transparent manner via worker pools. Internally, the observational data set is split into approximately equal-sized chunks of observations and the likelihood function is then calculated for each chunk separately via the map function of the pool object. To enable parallel processing, the user simply has to provide an appropriate worker pool to the \verb|p()| function of the likelihood instance.

The current version of \leopy{} works well with the multiprocessing and MPI pools provided by the `schwimmbad' module. For instance, the following code parallelizes the minimal example of section \ref{sect:minimal example} with the help of a multiprocessing pool.

\begin{lstlisting}[language=Python]
import leopy
from schwimmbad import MultiPool
pool = MultiPool()

d = {'v0': [1, 2], 'e_v0': [0.1, 0.2],
     'v1': [3, 4], 'e_v1': [0.1, 0.1]}
obs = leopy.Observation(d, 'testdata')
    
like = leopy.Likelihood(obs, 
    p_true='gamma', p_cond='norm')
    
like.p([0.5, 0.7], [1, 2], 
       shape_true=[1.4, 2], pool=pool)
pool.close()
\end{lstlisting}

For parallelization via MPI, the following lines from the code above
\begin{lstlisting}[language=Python]
from schwimmbad import MultiPool
pool = MultiPool()
\end{lstlisting}
need to be replaced with
\begin{lstlisting}[language=Python]
from schwimmbad import MPIPool
pool = MPIPool()
if not pool.is_master():
    pool.wait()
    import sys
    sys.exit(0)
\end{lstlisting}
This python script can then be executed like any other MPI program on the given platform, e.g., via 
\lstset{emph={processes, code, py}, emphstyle=\itshape}
\begin{lstlisting}[language=bash]
mpirun -n processes python code.py
\end{lstlisting}
from the command line.

\section{\leopy: The star forming sequence of nearby galaxies}
\label{sect:Application}

The method outlined in the previous section is applicable for a range of problems that involve the analysis of observational data.  To exemplify the use of \leopy{} on an astrophysical problem, I will use data from the xGASS \citep{Catinella2018} survey to study the properties of the so-called star forming sequence of galaxies \citep{Daddi2007a, Noeske2007d, Elbaz2007, Iyer2018}, i.e., the correlation between star formation rates (SFRs) and stellar masses ($M_{\rm star}$) of galaxies. 

The xGASS catalog contains the properties of $\sim{}10^3$ nearby galaxies ($0.01\leq{}z\leq{}0.05$) over a  large range of stellar masses ($\sim{}10^9-10^{11.5}\,M_\odot$) and with  robustly estimated SFRs and uncertainties \citep{Janowiecki2017}. xGASS is a follow-up survey of the GALEX Arecibo SDSS Survey (GASS, \citealt{Catinella2010}).

In this section, I will characterize the SFR distribution of galaxies in xGASS with the help of \leopy{} and compare these findings with the literature. Section \ref{sect:DataSet} briefly describes the data set. Section \ref{sect:ModelingSFS} details how the likelihood for galaxies in xGASS is calculated for given model parameters. Finally, section \ref{sect:FittingSFS} presents the outcome of the likelihood maximization and compares these findings with published work.

\subsection{The data set}
\label{sect:DataSet}

The following analysis is based on the `representative sample' from xGASS (available at \url{http://xgass.icrar.org}). The data set contains various properties (notably SFRs, stellar masses, $H_{\rm I}$ masses, their measurement uncertainties and detection limits) of 1179 nearby galaxies. In the present analysis, the only properties of interest are the SFRs, their uncertainties, and the stellar masses.

Of the 1179 xGASS galaxies, 817 galaxies have UV or IR coverage and reliable uncertainty estimates \citep{Janowiecki2017}. In contrast, the SFRs of 362 galaxies have been estimated with a different method \citep{Wang2011a} and no uncertainties are provided. In order to simplify the analysis, I mark the SFRs and uncertainties of the latter subset of galaxies as missing.

Many galaxies in xGASS have SFRs much lower than their reported uncertainties, yet no SFRs are reported to be negative. This implies that either the SFR errors (at low SFRs) are not normally distributed (e.g., follow a Poisson distribution) or that (low) SFRs are biased. To mitigate this problem, I censor all SFR values that are lower than $f_{\rm d}$ times the reported SFR uncertainties, i.e., the detection limit of such censored SFRs is given by $f_{\rm d}$ times their uncertainties. For the likelihood maximization (see below), I adopt $f_{\rm d}=1$, but choosing, e.g., $f_{\rm d}=2$, results in qualitatively similar predictions.

A different problem arises due to the potential contribution from a small fraction of 'star-bursting' galaxies (e.g., \citealt{Sargent2012}), i.e., from galaxies that lie well above the star forming sequence.  Their extreme positions in the SFR -- stellar mass plane combined with their quoted small SFR errors could potentially bias the maximum likelihood analysis. I model such galaxies as an additional, lognormal component, following the approach outlined in section \ref{sect:Outliers}.

xGASS does not provide (statistical) uncertainties of the stellar masses. In this analysis, I assume that measurements of $M_{\rm star}$ are error-free. I checked that adopting a small but constant uncertainty (e.g., 0.06 dex, i.e. 15\%) changes none of the results in a significant manner. It increases, however, the computation time as the likelihood computation involves an additional marginalization over the (unknown) true stellar mass.

The xGASS data with the modifications described above can be found in the code repository of \leopy{} in form of a CSV file.

\subsection{Modeling the star forming sequence}
\label{sect:ModelingSFS}

I model the SFR distribution at fixed stellar mass as a \emph{zero-inflated} gamma distribution following \cite{Feldmann2017} with the addition of a small lognormal component for starbursting galaxies. The distribution parameters are $\theta=(f_{\rm zero}, a, b, \phi)$, where $f_{\rm zero}$ is the fraction of galaxies with ${\rm SFR}=0$, $a$ and $b$ are the shape and scale parameters of the gamma distribution that describes the distribution of SFRs in the non-zero-inflated component, and $\phi$ is a set of three parameters that describe the position, width, and relative fraction of starbursting component. To simplify the presentation, I will not explicitly highlight $\phi$ in the subsequent discussion.

At a basic level, the zero-inflated component should be understood as a free parameter that accounts for deviations of the SFR distribution from a gamma distribution at very low SFRs. Whether galaxies in the zero-inflated component constitute a separate class of galaxies or whether they belong to an extended tail of the star forming sequence, would require reliable measurements of SFRs at extremely low levels. From a practical point of view, given realistic measurement uncertainties, these two alternatives may be indistinguishable \citep{Feldmann2017}.

For the regression analysis, I choose the logarithm of the stellar mass as the independent variable, $t = \lg{}M_{\rm star}$, and introduce new parameters $\omega$ that determine the scaling of the distribution parameters with $t$. Specifically, the logit\footnote{${\rm logit}(x)=\ln(x/(1-x))$} of $f_{\rm zero}$ is modeled as a linear function of $t$ with slope $m_{\rm zero}$ and intercept $n_{\rm zero}$. The logarithms of $a$ and $b$ are assumed to scale linearly with $t$ with slopes $m_a$, $m_b$ and intercepts $n_a$, $n_b$. These 6 parameters are the primary fitting parameters\footnote{The actual fitting uses slope angle ($\phi=\arctan(m)$) and perpendicular distance to the origin ($d=n\cos(\phi)$) as main parameters following the suggestion by \cite{Hogg2010}.}, i.e., $\omega=(m_{\rm zero}, n_{\rm zero}, m_a, n_a, m_b, n_b)$. Given that the stellar mass is assumed to be error free, the linear regression analysis described here is simpler than the approach outlined in section \ref{sect:Regression} as it does not involve a numerical integration over $t$.

I model the distribution of true SFRs as follows:
\begin{equation}
\label{eq:SFR_model}
\begin{split}
p({\rm SFR}\vert{}\,t, \omega) =\,& p({\rm SFR}\vert{}\,\theta(t, \omega))  \\
=\,& \left[1-f_{\rm zero}(t, \omega)\right]\,p_{\rm cont}({\rm SFR}\, \vert{}\, t, \omega) \\ 
&+  f_{\rm zero}(t, \omega)\, p_{\rm zero}({\rm SFR}).
\end{split}
\end{equation}
The probability density of the SFRs is a mixture of an extended probability density $p_{\rm cont}$ (assumed to be a gamma distribution) and a second, much more concentrated distribution $p_{\rm zero}$ (modeled by a halfnormal distribution with small scale parameter) representing the zero-component. As mentioned above, the mixture also may include a third component corresponding to starbursting galaxies.

Starting from the generative model provided by (\ref{eq:SFR_model}), \leopy{} computes, for each observation $i$, the following likelihood 
\[
p({\rm SFR}^{\rm obs}_i\,\vert\,t_i, \omega) = \int d{\rm SFR}\,p({\rm SFR}\vert{}\,t_i, \omega)\,p({\rm SFR}^{\rm obs}_i\,\vert\,{\rm SFR}, t_i).
\]
The full likelihood is then
\[
\mathcal{L} = \prod_i p({\rm SFR}^{\rm obs}_i\,\vert\,t_i, \omega).
\]
The above equations are modified as outlined in section \ref{sect:CensoredAndMissingData} to account for missing data and data censoring. 
To find the parameters that maximize the full likelihood, I employ the basinhopping routine and the SLSQP minimizer that are part of the scipy.optimize package.

\begin{figure}[t]
\includegraphics[width=85mm]{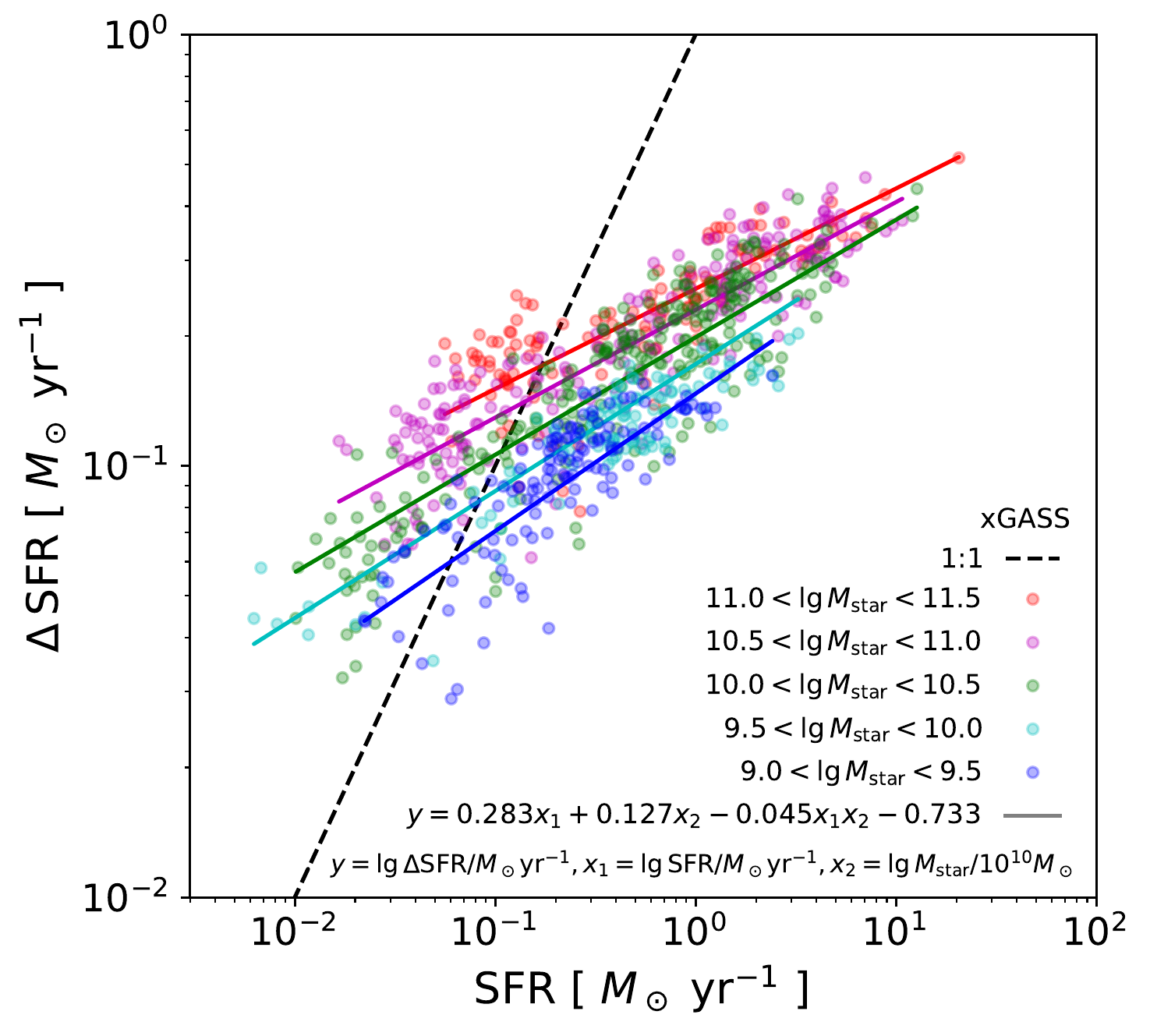}
\caption{SFRs and their measurement uncertainties ($\Delta{}{\rm SFR}$) in the xGASS sample. A multiple linear regression with $\lg{}\Delta{}{\rm SFR}$ as the dependent variable and $\lg{}{\rm SFR}$ and $\lg{}M_{\rm star}$ as independent variables 
 is able to recover the main trends. The regression model includes an interaction term (see figure legend). SFRs below $\sim{}0.1$ $M_\odot$ yr$^{-1}$ are highly affected by measurement errors.}
\label{fig:SFR_uncertainty}
\end{figure}

I assess the quality of the maximum likelihood regression with the help of mock samples. To this end, I first randomly draw $N_{\rm mock}$ stellar masses from the actual mass distribution of xGASS. Next, I use the maximum likelihood parameters to compute, for each of the $N_{\rm mock}$ mock objects, the SFR distribution and draw a random variate from it. The so generated SFRs are mock versions of the `true' (latent) SFRs. In order to properly compare their distribution with the observed distribution in xGASS, I further need to add realistic SFR errors and to censor mock objects with low observed SFRs.

I calculate the average SFR uncertainty for galaxies of a given mass and SFR via multiple linear regression using the 817 galaxies in xGASS with reasonable SFR uncertainties. Specifically, by using $\lg{}{\rm SFR}$ and $\lg{}M_{\rm star}$ as predictors, I obtain
\begin{multline}
\label{eq:SF_uncertainty}
\lg \frac{\Delta{}{\rm SFR}}{M_\odot\,{\rm yr}^{-1}} = -0.733 + 0.283 \lg \frac{\rm SFR}{M_\odot\,{\rm yr}^{-1}} \\ + 0.127 \lg\frac{M_{\rm star}}{10^{10} M_\odot}
- 0.045 \lg \frac{\rm SFR}{M_\odot\,{\rm yr}^{-1}} \lg \frac{M_{\rm star}}{10^{10} M_\odot},
\end{multline}
see Fig.~\ref{fig:SFR_uncertainty}. To mimic observed SFRs in xGASS, I thus add to each mock SFR a random variate drawn from a normal distribution with zero mean and with a standard deviation given by the above SFR uncertainty. Finally, mock SFRs are censored if they are lower than $f_{\rm d}$ times their SFR uncertainty.

\subsection{Properties of the star forming sequence in low redshift galaxies}
\label{sect:FittingSFS}

\begin{figure}[t]
\includegraphics[width=85mm]{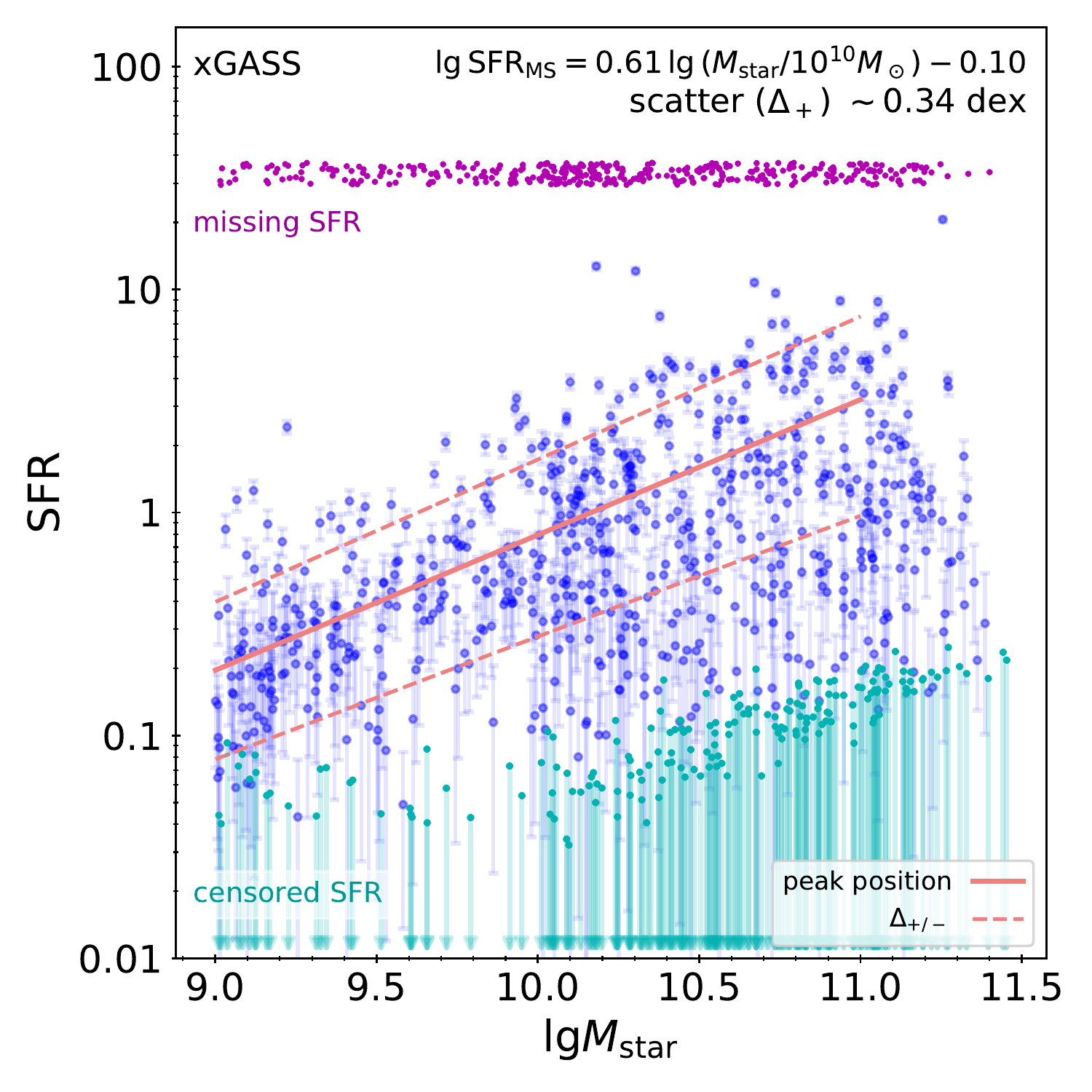}
\caption{SFR and stellar masses of low-redshift galaxies ($0.01<z<0.05$) reported by the xGASS survey \citep{Catinella2018}. Error bars show the measurement uncertainties of SFRs in xGASS. SFR estimates without reliable uncertainty estimates are marked as missing data and are shown as small circles near the top of the figure. In contrast, SFRs with reliable uncertainty estimates but with values smaller than their uncertainties are censored and indicated by the downward pointing arrows.
The peak position of the star forming sequence is shown by the solid line. It is derived from a maximum likelihood analysis (see text) using \emph{all} galaxies in the sample with $9\leq{}\lg{}M_{\rm star}/M_\odot\leq{}11$. SFRs at fixed $M_{\rm star}$ are modeled as a zero-inflated gamma distribution with a small additional lognormal component representing starbursting galaxies. 
The star forming sequence scales sub-linearly with stellar mass, ${\rm SFR}\propto{}M_{\rm star}^{0.61}$, in good agreement with estimates by \cite{Catinella2018} based on a different method. Dashed lines show the upward and downward scatter of the star forming sequence for a given stellar mass as defined in equation (\ref{eq:scatter_def}) with $r=1$. The upward scatter ($\Delta{}_+$) of the star forming sequence varies between 0.31 dex at $M_{\rm star}=10^9$ $M_\odot$ and 0.37 dex at $M_{\rm star}=10^{11}$ $M_\odot$. The downward scatter ($\Delta{}_-$) is significantly larger and varies more strongly with mass.}
\label{fig:SFsequence}
\end{figure}

\begin{figure*}[t]
\begin{tabular}{cc}
\includegraphics[width=85mm]{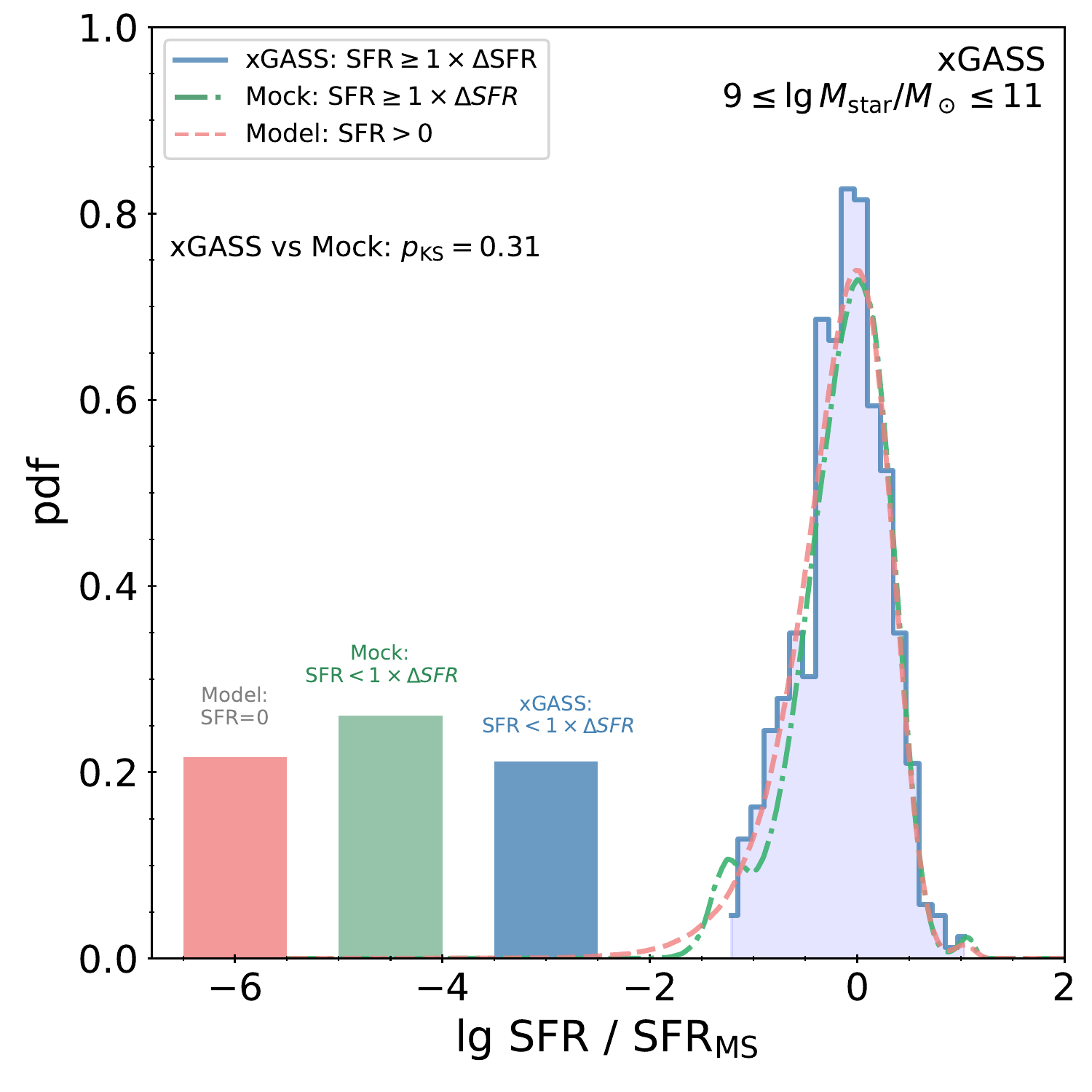} &
\includegraphics[width=85mm]{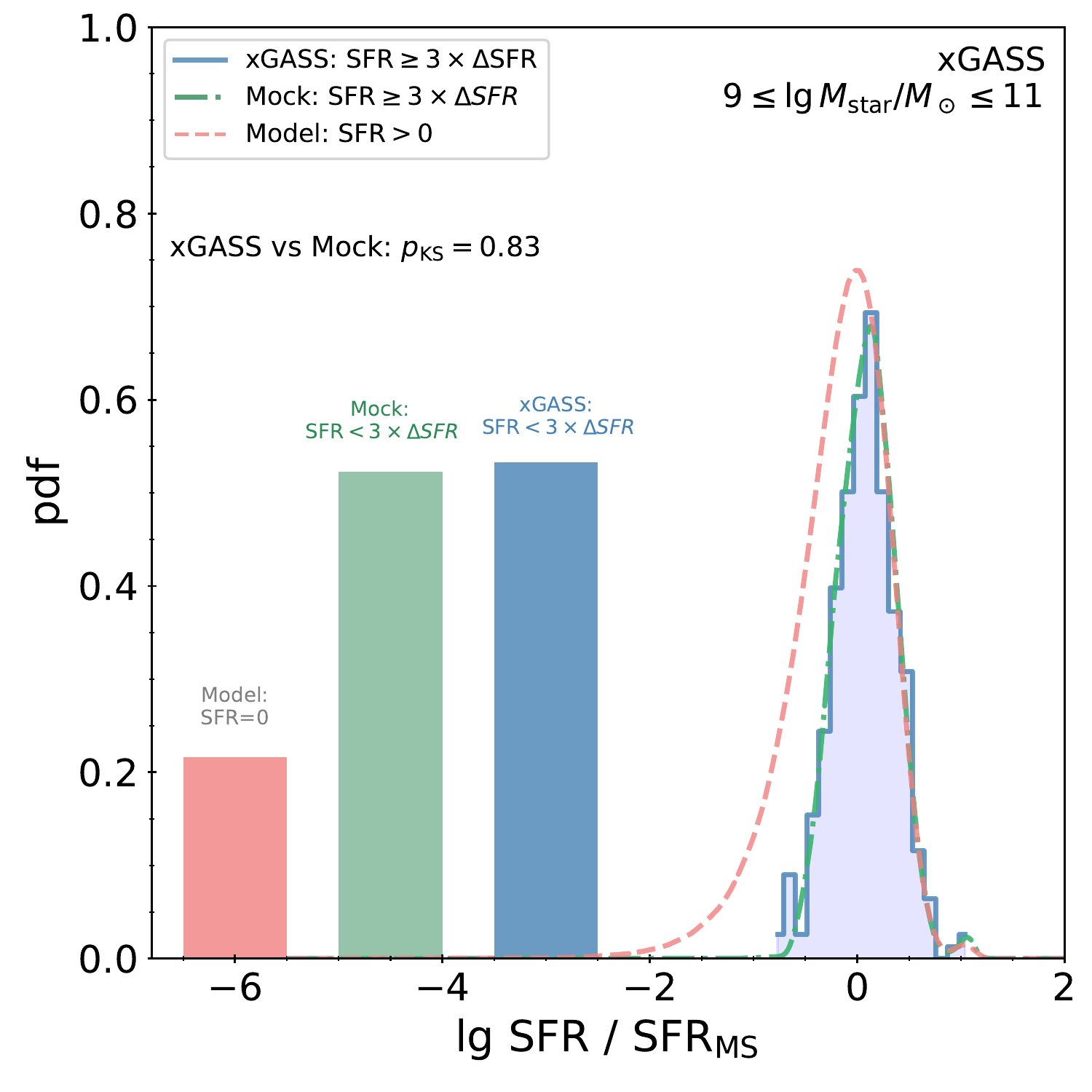}
\end{tabular}
\caption{Distribution of SFRs relative to the peak position of the star forming sequence and fraction of galaxies with SFRs consistent with being zero. The left panel (right panel) considers SFR with values smaller than $f_{\rm d}=1$ ($f_{\rm d}=3$) times their uncertainties as being consistent with zero.
In each panel, the histogram shows the distribution of the observed SFRs in xGASS for galaxies with ${\rm SFR}\geq{}f_{\rm d}\,\Delta{}{\rm SFR}$ and $10^{9}$ $M_\odot\leq{}M_{\rm star}\leq{}10^{11}$ $M_\odot$. This distribution is compared with the SFR distributions of a mock sample ($N_{\rm mock}=10^6$) created as described in the text for the same stellar mass range. Mock SFRs with ${\rm SFR}\geq{}f_{\rm d}\,\Delta{}{\rm SFR}$ are shown by the dot-dashed line. These SFRs include measurement errors and they well approximate the distribution of observed SFRs with ${\rm SFR}\geq{}f_{\rm d}\,\Delta{}{\rm SFR}$ in the xGASS sample (solid line and histogram). The distribution of the \emph{true} SFRs in the mock sample with ${\rm SFR}>0$ is shown by the dashed line. The latter distribution is highly asymmetrical with an extended tail towards low SFRs. This suggests that galaxies with SFRs one or two orders of magnitude below the peak position may be part of the star forming sequence.}
\label{fig:SFdistribution}
\end{figure*}

Fig.~\ref{fig:SFsequence} shows SFRs of the galaxies in the xGASS sample as a function of their stellar mass. Slope and scatter of this star forming sequence place important constraints on galaxy formation models \citep{Iyer2018, Donnari2019, Caplar2019} and may help to uncover the drivers of star formation activity in galaxies.

SFRs at fixed stellar mass show a large degree of scatter but, at least their upper limits, appear to increase with stellar mass. A significant fraction of galaxies in the present sample have either missing SFRs (those without reliable UV and IR coverage) or have censored SFRs (those with SFRs lower than their uncertainties). A proper analysis of this data set thus requires a careful modeling of missing data, data censoring, and measurement uncertainties.

SFRs of galaxies belonging to the star forming sequence are usually assumed to be log-normally distributed (e.g., \citealt{Guo2013c, Davies2018}). Given that the distribution of $\lg{\rm SFR}$ is asymmetric, see Fig.~\ref{fig:SFsequence}, galaxies with low SFRs are often excluded when constraining the parameters of the star forming sequence. The resulting approaches to fit the parameters of the star forming sequence are consequently somewhat ad hoc.

In contrast, the approach outlined in section \ref{sect:ModelingSFS} does not separate galaxies into different classes based on their star formation activity. Instead, all galaxies of a given mass range are included in the analysis. The asymmetry of the $\lg{\rm SFR}$ distribution is accounted for by adopting a (zero-inflated) gamma distribution (with shape and scale parameters $a$ and $b$) instead of a log-normal distribution of the SFRs. A (small) starburst component can be explicitly added as a separate component if required.

The peak position of the star forming sequence is naturally defined as the SFR that corresponds to the dominant mode of the log-transformed SFR. Given a gamma distributed SFR with shape and scale parameters $a$ and $b$, i.e., ${\rm SFR}\sim{}\Gamma(a, b)$, $\ln{}{\rm SFR}$ is exp-gamma distributed and has a mode at $\ln(ab)$ \citep{SAS2018}. The peak of the star forming sequence is thus at $ab$. Furthermore, $ab$ is the mean of the gamma distributed SFR.

Hence, the peak position of the star forming sequence, ${\rm SFR}_{\rm MS}$, is at 
\begin{equation}
\label{eq:peakSFS}
{\rm SFR}_{\rm MS} = \langle{}{\rm SFR}\rangle{}_{+} = ab,
\end{equation}
where  $\langle{}{\rm SFR}\rangle{}_{+}$ is the average SFR of all galaxies in the gamma-distributed component.

This intriguing result does not necessarily hold for a generic SFR distribution. For instance, if SFRs were lognormally distributed with shape $s$ and scale $\sigma$ (see caption of Table.~\ref{tab:example}), then the peak position of the star forming sequence would be at $\sigma$, while the average SFR would be $\sigma \exp{(s^2/2)}$.
  
The scatter of the star forming sequence is often defined in a symmetric fashion either as the standard-deviation of the $\lg{}{\rm SFR}$ distribution or via its percentiles (e.g., \citealt{Elbaz2007, Guo2013c, Donnari2019}). However, the following alternative better accounts for the asymmetry of the $\lg{}{\rm SFR}$ distribution. Let $p_{\rm peak}$ be the probability density of the (non-zero) mode of the $\lg{}{\rm SFR}$ distribution and $r>0$ be a chosen real number. The upward (downward) scatter $\Delta{}_+$ ($\Delta{}_-$) is defined as $1/r$ times the smallest increase (decrease) in $\lg{}{\rm SFR}$ that reduces the probability density from $p_{\rm peak}$ to $e^{-r^2/2}p_{\rm peak}$. The factor $e^{-r^2/2}$ ensures that $\Delta{}_+=\Delta{}_-=\sigma$ for a normal distribution with a standard deviation of $\sigma$ for any chosen $r>0$. For general non-normal distributions, however, $\Delta{}_+$ and $\Delta{}_-$ differ from each other and depend on the choice of $r$. 

It is easy to show that the upward and downward scatter of the logarithm of a gamma distributed variable with given shape $a>0$ and (arbitrary) scale $b>0$ parameters is
\begin{equation}
\label{eq:scatter_def}
\Delta{}_{\pm}^{(r)} = \pm\frac{1}{r}\lg\left(-W_{\mp}\left(- e^{-\frac{r^2}{2a}}/e\right)\right)
\end{equation}
where $W_+$ and $W_-$ are the principal and the -1 branch of the Lambert $W$ function. In this section, I adopt the above definition with $r=1$. In simple terms, $\Delta{}_{\pm}$ measures the upward and downward scatter of a gamma distributed variable (e.g., the SFR) in \emph{dex}.

The most likely value of the parameters $\omega$ of the SFR distribution can be inferred from a maximum likelihood analysis. For instance, $\omega_{\rm ML} = (1.755, -1.771, -0.099,  0.097, 0.706, -0.198)$ if all xGASS galaxies with $10^{9}$ $M_\odot\leq{}M_{\rm star}\leq{}10^{11}$ $M_\odot$ are included in the analysis.

\begin{table*}
\begin{center}
\begin{tabular}{llll|llllllll}
\hline\hline
$\lg M_{\rm star} / M_\odot$ & $N_{\rm gal}$  &  $N_{\rm cen}$ & $N_{\rm mis}$ & $m_{\rm peak}$ & $n_{\rm peak}$ & $m_{\Delta{}_+}$ & $n_{\Delta{}_+}$ & $m_{\Delta{}_+^{(2)}}$ & $n_{\Delta{}_+^{(2)}}$ & $m_{\rm zero}$ & $n_{\rm zero}$ \rule[-1.5ex]{0pt}{0pt} \\ \hline 
9 -- 10 & 343 & 26 & 107 & 0.71 & -0.183 & -0.028 & 0.17 & -0.025 & 0.16 & 0.94 & -1.51 \\
9 -- 10.5 & 707 & 76 & 233 & 0.71 & -0.084 & 0.041 & 0.33 & 0.033 & 0.29 & 3.29 & -1.92 \\
9 -- 11 & 1011 & 146 & 322 & 0.61 & -0.100 & 0.033 & 0.34 & 0.026 & 0.30 & 1.76 & -1.77 \\
9 -- 11.5 & 1179 & 184 & 362 & 0.52 & -0.131 & 0.068 & 0.36 & 0.053 & 0.31 & 1.18 & -1.85 \\
\hline\hline
\end{tabular}
\caption{Properties of the star forming sequence based on an analysis of the xGASS sample with \leopy{}, see section \ref{sect:FittingSFS}. The first column shows the stellar mass range of the galaxies included in the maximum likelihood analysis. Columns 2-4 list the total number of galaxies included in the analysis ($N_{\rm gal}$), the number of galaxies with censored SFRs, and the number of galaxies with missing SFRs. Columns 5 and 6 provide the most likely values of the slope and intercept of the peak position of the star forming sequence, i.e., $\lg{}{\rm SFR} = m_{\rm peak}  \lg (M_{\rm star}/10^{10} M_\odot )  + n_{\rm peak}$. Columns 7-10 contain the scaling parameters of the upward scatter $\Delta{}_+ = m_{\Delta{}_+} \lg (M_{\rm star}/10^{10} M_\odot )  + n_{\rm \Delta{}_+} $ of the star forming sequence with $r=1$ (columns 7 and 8) and $r=2$ (columns 9 and 10), see text. The final two columns provide the parameters of the zero-inflated component, $f_{\rm zero}=e^\xi/(1+e^\xi)$ with $\xi = m_{\rm zero} \lg (M_{\rm star}/10^{10} M_\odot) + n_{\rm zero}$. 
SFRs increase (sub-linearly) with increasing stellar mass. The upward scatter is about $0.3-0.35$ dex for galaxies with $M_{\rm star}\sim{}10^{10}$ $M_\odot$ and it increases with increasing mass. The zero-inflated component is sub-dominant in low mass galaxies (typically $f_{\rm zero}<20\%$ for $M_{\rm star}\leq{}10^{10}$ $M_\odot$) and a major component at high masses ($f_{\rm zero}\gtrsim{}50\%$ for $M_{\rm star}>10^{11}$ $M_\odot$).}
\label{tab:SFS}
\end{center}
\end{table*}

The peak position and upward scatter of the star forming sequence are directly related to the shape and scale parameters $a$ and $b$ of the gamma distribution. According to equation (\ref{eq:peakSFS}), $m_{\rm peak}=m_{\rm a} + m_{\rm b}$ and $n_{\rm peak}=n_{\rm a} + n_{\rm b}$ are the slope and intercept of the common logarithm of the peak position of the star forming sequence. The upward scatter is a somewhat complex function of the shape parameter $a$, see equation (\ref{eq:scatter_def}), but its scaling with $\lg{}M_{\rm star}$ is close to linear. Hence, it is possible to retrieve the slope ($m_{\rm \Delta{}_+}$) and intercept ($n_{\rm \Delta{}_+}$) of the upward scatter from a simple linear regression of $\Delta{}_{+}$ as a function of  $\lg{} M_{\rm star}$.

Table~\ref{tab:SFS} summarizes the maximum likelihood parameters of the star forming sequence for various stellar mass ranges. E.g., for $10^{9}$ $M_\odot\leq{}M_{\rm star}\leq{}10^{11}$ $M_\odot$:
\begin{equation}
\label{eq:SFR_MS_scaling}
\begin{split}
f_{\rm zero} =& \frac{e^{\xi}}{1+e^\xi}\textrm{ with }\xi=1.76 \lg \frac{M_{\rm star}}{10^{10} M_\odot} - 1.77,\\
\lg {\rm SFR}_{\rm MS} =&\, 0.61 \lg \frac{M_{\rm star}}{10^{10} M_\odot} - 0.10, \\
\Delta{}_{+} =&\, 0.033 \lg{} \frac{M_{\rm star}}{10^{10} M_\odot} + 0.34.
\end{split}
\end{equation}

The regression result depends on the mass range of galaxies included in the fit. In particular, restricting the analysis to lower masses increases the slope of the peak position and decreases slope and normalization of the scatter. The former finding is qualitatively consistent with a flattening or turning over of the slope of the star forming sequences at high masses \citep{Whitaker2014b, Eales2018}. 

The decrease of the scatter with decreasing stellar mass for $9<M_{\rm star}/M_\odot<11$ is consistent with the observational literature (e.g., \citealt{Guo2013c, Davies2018}, but cf. \citealt{Bauer2013}). Such a scaling has also been qualitatively reproduced by some numerical simulations (e.g., \citealt{Sparre2015b, Donnari2019} but cf. \citealt{Furlong2015a}).

The precise value of the scatter of the star forming sequence is expected to depend on the observational technique used to infer SFRs (e.g., \citealt{Hopkins2014}). In particular, the scatter should decrease with increasing averaging time of the observational tracer due to temporal correlations of SFRs (e.g., \citealt{Caplar2019}). In line with these expectations, \cite{Davies2018} find that SFRs based on UV + IR observations show a lower scatter than those based on $H\alpha$ line measurements. Hence, an analysis using $H\alpha$ as SFR indicator would likely return a somewhat larger scatter than the one reported in Table~\ref{tab:SFS}.

 \cite{Catinella2018} estimate the scaling properties of the star forming sequence in the xGASS sample using a very different approach as the one outlined above. They find $\lg {\rm SFR}_{\rm MS} = 0.656 \lg (M_{\rm star} / 10^{10} M_\odot) - 0.166$ and a scaling of the upward scatter $0.088 \lg (M_{\rm star} / 10^{10} M_\odot ) + 0.276$. The results reported by \cite{Catinella2018} are similar to the ones reported in equation (\ref{eq:SFR_MS_scaling}). However, it should be noted that \cite{Catinella2018} estimate the peak position of the star forming sequence based on the subset of galaxies with $10^{9}$ $M_\odot\leq{}M_{\rm star}\leq{}10^{10}$ $M_\odot$. For this lower mass range, Table~\ref{tab:SFS} reports a steeper scaling of the peak position ($m_{\rm peak}=0.71$) and a smaller scatter ($\Delta{}_{+}=0.17$). 

The final task is to test how well the proposed model, a zero-inflated gamma distribution with parameters $\omega_{\rm ML}$, fits the observed distribution of SFRs. To explore this important question, I create mock observations that match the stellar mass distribution of the xGASS sample, see section \ref{sect:ModelingSFS}.

Fig.~\ref{fig:SFdistribution} shows the distribution of SFRs relative to the peak position of the star forming sequence as well as the fraction of galaxies with SFRs consistent with being zero. The latter are defined as galaxies with measured SFR smaller than the SFR uncertainties (left panel) and smaller than 3 times the SFR uncertainties (right panel).

The peak position of the star forming sequence is obtained via equation (\ref{eq:peakSFS}) with $a$ and $b$ provided by the regression analysis with $f_{\rm d}=1$. Mock data based on the zero-inflated gamma distribution model reproduce the shape of the observed SFR distributions in xGASS sample quite well. The visually good agreement is quantified by a KS test on the non-zero component, i.e., after excluding galaxies with SFRs smaller than $f_{\rm d}=1$ or $f_{\rm d}=3$ times their uncertainties. In either case, a KS test cannot reject the null hypothesis that the observed SFRs in xGASS are drawn from the same distribution as the mock data at a significance level of 0.05.

Interestingly, the distribution of the \emph{true} SFRs in the mock sample is highly asymmetrical with an extended tail towards low SFRs. This suggests that galaxies with SFRs one or two orders of magnitude below the peak position may be part of the star forming sequence. This result is in line with the ideas proposed by \cite{Feldmann2017} and \cite{Eales2017a} that star forming and quiescent galaxies are not intrinsically different classes of galaxies but rather form a continuous sequence. Further evidence for the asymmetric shape of the true SFR distribution is provided by numerical simulations that find that (zero-inflated) gamma distributions are a good match to the SFR distributions \citep{Feldmann2017, Donnari2019}.

The model also predicts a significant fraction of galaxies with extremely low or vanishing SFRs (the zero component in the model). Depending on the exact shape of the distribution of these extremely low SFRs, the distribution of $\lg{}{\rm SFR}$ may or may not contain a second mode at such low SFRs.

To summarize, a zero-inflated gamma distribution with parameters that depend on stellar mass provides a reasonable model for the distribution of the actual SFRs in nearby galaxies. A maximum likelihood analysis allows to extract the relevant model parameters based on the measured SFRs of all galaxies in a given stellar mass range, i.e., the approach outlined in this section does not require ad hoc procedures for removing the large number of galaxies with low SFRs to fit the slope, intercept, and scatter of the star forming sequence. It can also be easily adapted to account for additional components, such as starbursting galaxies. Furthermore, the results from the maximum likelihood analysis are in approximate agreement with the findings by \cite{Catinella2018} using a different methodology. Mock data created via the zero-inflated gamma distribution model reproduce well the shape of the observed SFR distribution.

\section{Summary and conclusion}
\label{sect:Summary}

Statistical inference is a central pillar of data science and statistics. Applying statistical inference techniques, such as regression, to real-world data is often hampered by a number of challenges. Such data sets may feature missing values, censored data entries, data correlations, outliers, and (potentially correlated) measurement errors. Unless handled appropriately, such data can result in biased or inconsistent parameter estimates, see section \ref{sect:Motivation}.

The approach presented in this paper extends previous work (e.g., \citealt{Kelly2007, Hogg2010, Robotham2015}) by properly accounting for data correlations in joint distributions with \emph{user-defined marginal distributions} as well as by being able to handle censored data with both lower and upper limits. Section \ref{sect:Method} outlines the main approach of dealing with such data sets in the context of statistical inference.

Specifically, data correlations are handled by Gaussian copulas (\ref{sect:corrobs}). Correlated measurement errors are properly propagated to obtain the joint distribution of measured values (\ref{sect:S}). Censored data and missing data in the `missing at random` approximation are handled consistently by partially integrating the joint probability density function (\ref{sect:CensoredAndMissingData}). Finally, outliers are easily accounted for by adding additional mixture components to the input marginal distributions (\ref{sect:Outliers}).

To exemplify the usefulness of the outlined method, I re-analyzed the publicly available xGASS data set \citep{Catinella2018} to study the properties of the star forming sequence of nearby galaxies. In contrast with previous approach in the literature, galaxies are not ad hoc excluded from the analysis based on their relative position to the star forming sequence. The distribution of the star formation rates (SFRs) at fixed stellar mass ($M_{\rm star}$) is modeled as a zero-inflated gamma distribution as suggested by previous work \citep{Feldmann2017}. A maximum likelihood determination of the distribution parameters using all galaxies with $9\leq{}\lg{}M_{\rm star}/M_\odot\leq{}11$ resulted in a model that matches the observed SFR distribution in xGASS extremely well. This model has the following properties.
\begin{itemize}
\item In qualitative agreement with the literature, I find that the peak position of the star forming sequence has a positive, but sub-linear, trend with stellar mass. The slope decreases if more massive galaxies are included indicative of a turn-over of the star forming sequence at the massive end.
\item The intrinsic upward scatter of the star forming sequence is about 0.34 dex for galaxies with $M_{\rm star}=10^{10}$ $M_\odot$ and varies between 0.31 dex at $M_{\rm star}=10^{9}$ $M_\odot$ and 0.37 dex at $M_{\rm star}=10^{11}$ $M_\odot$.
\item The intrinsic downward scatter is significantly larger than the upward scatter which is a consequence of the $\lg{}{\rm SFR}$ distribution being asymmetric with an extended tail towards low SFRs. Galaxies with SFRs up to two orders of magnitude below the peak position may thus be part of the star forming sequence.
\item About 15\% (3\%, 49\%) of the galaxies with $M_{\rm star}=10^{10}$ $M_\odot$ ($M_{\rm star}=10^{9}$ $M_\odot$, $M_{\rm star}=10^{11}$ $M_\odot$) have very low ($\lesssim{}0.01\times{\rm SFR}_{\rm MS}$) or, perhaps, vanishing SFRs. The strong mass trend of galaxies with such low SFRs may be reflective of the prevalence of `red and dead' galaxies at the massive end.
\end{itemize}

The proposed methodology has been fully implemented as an open source Python package. This `Likelihood Estimation for Observational data with Python` (\leopy{}) code comes with an extensive testing suite and with a number of example problems. Aside from schematic Figs.~\ref{fig:flowchart} and \ref{fig:multilevel}, all figures in this paper can be recreated with Python scripts that are provided as part of the package. \leopy{} is easy-to-use and, combined with minimization or Monte Carlo methods, can be applied to a variety of different data analysis problems. I hope its availability will facilitate the analysis of observational data with censored, missing, and correlated values.

\section*{Acknowledgements}
The author would like to thank the referee for suggestions that helped to improve the quality of the paper. Furthermore, the author would like to thank Leonhard Held and Reinhard Furrer for fruitful discussions during an early stage of this work, and Romain Teyssier for commenting on the completed manuscript. The author is also grateful to Barbara Catinella for help with the xGASS data set. RF acknowledges financial support from the Swiss National Science Foundation (grant no 157591). This research has made use of NASA's Astrophysics Data System Bibliographic Services.






 
 
 \appendix 
 \section{Multivariate normal distribution with normally distributed errors.}

I argued in section \ref{sect:MultivariateNormal} that the likelihood is a multivariate normal distribution if both $\mathbf{Y}$ and  $\mathbf{Y}^{\rm obs}\vert{}\mathbf{Y}$ are normally distributed. In this paragraph I show explicitly that the approach outlined in this paper (based on Gaussian copulas) reproduces this result as a special case.
 
Specifically, let $f_j$ be a univariate, normal density with mean $\mu_j$ and variance $\sigma^2_j$ for all $j=1,\ldots{},K$, and $\tilde{f}^{\rm c}_j$ be a univariate, normal density with zero mean and variance $(\sigma^{\rm c}_j)^2$. The conditional probability is then $f^{\rm c}_j(y^{\rm obs}_j\,\vert\,y_j)=\tilde{f}^{\rm c}_j(y^{\rm obs}_j - y_j)$. In this case, 
\[
f^{\rm obs}_j(y^{\rm obs}) = \int dy_j\, f^{\rm c}_j(y^{\rm obs}_j \,\vert\,y_j)\,f_j(y_j) = \left(\tilde{f}^{\rm c}_j * f_j \right)(y^{\rm obs}),
\]
i.e., $f^{\rm obs}$ is the convolution of $\tilde{f}^{\rm c}$ and $f$ and, therefore, also a normal density. Its mean and variance are $\mu^{\rm obs}_j=\mu_j$ and $(\sigma^{\rm obs}_j)^2=(\sigma^{\rm c}_j)^2+\sigma_j^2$. The correlation between the true values ($y_j$) is given by a Gaussian copula with correlation matrix $R$, while the observational errors have the correlation matrix $R^{\rm c}$. It is instructive to evaluate equation (\ref{eq:fullLikelihood}) in this case.

The standardized variables are
\[
z_j = \frac{y_j - \mu_j}{\sigma_j},\textrm{ and }z^{\rm obs}_j = \frac{y^{\rm obs}_j - \mu_j}{\sqrt{(\sigma^{\rm c}_j)^2+\sigma_j^2}}.
\]
Choosing $y^+_j = y^{\rm obs}_j$, implies $z^+_j=(y^{\rm obs}_j-\mu_j)/\sigma_j$, $z^-_j=-z^+_j$, $y^-_j= 2\mu_j-y^{\rm obs}_j$, and
\[
a_j=\ln{}\frac{f^{\rm c}_j(y^{\rm obs}_j \,\vert\, y^+_j)}{f^{\rm c}_j(y^{\rm obs}_j \,\vert\, y^-_j)}=\ln{}\frac{\tilde{f}^{\rm c}_j(0)}{\tilde{f}^{\rm c}_j(2y^{\rm obs}_j - 2\mu_j)}=\frac{2(y^{\rm obs}_j-\mu_j)^2}{(\sigma^{\rm c}_j)^2}.
\]
Inserting the above into equation (\ref{eq:Tjj2}) shows that
\[
b_j = \frac{(\sigma^{\rm c}_j)^2}{2\,\sigma_j\sqrt{\sigma_j^2+(\sigma^{\rm c}_j)^2}}\geq{}0,
\]
and
\begin{equation}
T_{jj} = -b_{j} + \sqrt{b_{j}^2+1} = \left(\frac{\sigma_j^2}{\sigma_j^2 + (\sigma^{\rm c}_j)^2}\right)^{1/2}.
\end{equation}
Hence, according to equation (\ref{eq:Smatrix2}),
\begin{equation}
\begin{split}
S_{j_1j_2} = & R_{j_1j_2}\left(\frac{\sigma_{j_1}^2}{\sigma_{j_1}^2 + (\sigma^{\rm c}_{j_1})^2}\frac{\sigma_{j_2}^2}{\sigma_{j_2}^2 + (\sigma^{\rm c}_{j_2})^2}\right)^{1/2} \\
&+ R^{\rm c}_{j_1j_2}\left(\frac{(\sigma^{\rm c}_{j_1})^2}{\sigma_{j_1}^2 + (\sigma^{\rm c}_{j_1})^2}\frac{(\sigma^{\rm c}_{j_2})^2}{\sigma_{j_2}^2 + (\sigma^{\rm c}_{j_2})^2}\right)^{1/2}
\end{split}
\end{equation}
and the likelihood can be computed from equation (\ref{eq:fullLikelihood}). Specifically, if there are no missing or censored data, then the likelihood of the parameters $\mu_j$, $\sigma_j$, $R$, $\sigma^{\rm c}_{ij}$, and $R^{{\rm c}, (i)}$ for the given data set $y^{\rm obs}_{ij}$ (with $i=1,\ldots,N$ and $j=1,\ldots{},K$) is
\begin{equation}
\begin{split}
\mathcal{L} =& \prod_{i=1}^{N}\left(\phi_{S^{(i)}}(\mathbf{z}^{\rm obs}_i) \times{} \prod_{j=1}^{K} \frac{f^{\rm obs}_{ij}(y^{\rm obs}_{ij})}  {\phi(z^{\rm obs}_{ij})}\right)  \\
=& \prod_{i=1}^{N}\left(\phi_{S^{(i)}}\left(\frac{\mathbf{y}^{\rm obs}_i-\bm{\mu}}{\bm{\sigma}^{\rm obs}_i}\right) \times{} \prod_{j=1}^{K} \frac{\phi_{(\sigma^{\rm obs}_{ij})^2}(y^{\rm obs}_{ij}-\mu_j)}  {\phi((y^{\rm obs}_{ij}-\mu_j)/\sigma^{\rm obs}_{ij})}\right) \\
=& \prod_{i=1}^{N}\left(\phi_{S^{(i)}}\left(\frac{\mathbf{y}^{\rm obs}_i-\bm{\mu}}{\bm{\sigma}^{\rm obs}_i}\right) \times{}\prod_{j=1}^{K}\frac{1}{\sigma^{\rm obs}_{ij}}\right) \\
=& \prod_{i=1}^{N}\phi_{W^{(i)}}\left(\mathbf{y}^{\rm obs}_i-\bm{\mu}\right).
\end{split}
\end{equation}
As expected, the likelihood is the product of a multivariate normal distribution with $W^{(i)}$ given by
\begin{equation}
\begin{split}
W^{(i)}_{j_1j_2} &= R_{j_1j_2} \sigma_{j_1}\sigma_{j_2} + R^{{\rm c}, (i)}_{j_1j_2} \sigma^{\rm c}_{ij_1}\sigma^{\rm c}_{ij_2},
\end{split}
\end{equation}
i.e., 
\[
W^{(i)} = \Sigma + \Sigma^{{\rm c}, (i)},
\]
where $\Sigma$ is the covariance matrix of the true data random variable $\mathbf{Y}$ and $\Sigma^{{\rm c}, (i)}$ is the covariance matrix of the measurement errors.

\end{document}